%% file: main_arxiv.tex
\documentclass[11pt]{article}

\usepackage[utf8]{inputenc}
\usepackage[T1]{fontenc}

\usepackage{geometry}
\usepackage{amsmath,amssymb,amsthm}

\usepackage{graphicx}
\usepackage{booktabs}  
\usepackage{multirow}
\usepackage{subcaption}  

\usepackage{algorithm}
\usepackage{algorithmic}

\usepackage{xcolor}
\usepackage[colorlinks=true,
            linkcolor=blue,
            citecolor=blue,
            urlcolor=blue]{hyperref}

\usepackage{cite}  
\usepackage{url}   
\usepackage{enumitem}  

\title{Beyond Discrete Categories: Multi-Task Valence-Arousal Modeling for Pet Vocalization Analysis}

\author{
  Junyao Huang\thanks{Corresponding author: huangjunyao@zhizibianjie.com}, Rumin Situ\\
  OmniEdge AI Consulting Co., Ltd.\\
  Shenzhen, China \\
  \texttt{huangjunyao@zhizibianjie.com, situruimin@zhizibianjie.com}
}

\date{\today}

\begin{document}

\maketitle

\begin{abstract}
\input{sections_arxiv/abstract}
\end{abstract}


\section{Introduction}
\label{sec:intro}
\input{sections_arxiv/introduction}

\section{Related Work}
\label{sec:related}
\input{sections_arxiv/related_work}

\section{Methodology}
\label{sec:method}
\input{sections_arxiv/methodology}

\section{Experiments}
\label{sec:experiments}
\input{sections_arxiv/experiments}

\section{Results and Analysis}
\label{sec:results}
\input{sections_arxiv/results}

\section{Discussion}
\label{sec:discussion}
\input{sections_arxiv/discussion}

\section{Conclusion}
\label{sec:conclusion}
\input{sections_arxiv/conclusion}



\input{main_arxiv.bbl}
% ...

\end{document}

%% file: sections_arxiv/abstract.tex

Pet emotion recognition from vocalizations has traditionally relied on discrete classification methods, which suffer from boundary ambiguity between adjacent categories and fail to capture subtle intensity variations---limitations that become particularly problematic in real-world applications such as pet welfare monitoring. We propose a continuous Valence-Arousal (VA) modeling approach that represents emotions in a two-dimensional space, combined with an automatic VA label generation algorithm that derives Arousal from RMS energy and Valence from spectral features (centroid, zero-crossing rate) augmented with emotion-specific priors, enabling large-scale annotation of 42,553 pet vocalization samples without requiring subjective self-reports. Our multi-task learning framework jointly trains VA regression alongside auxiliary classification tasks (emotion, body size, gender), leveraging knowledge transfer to enhance prediction: size classification forces learning of frequency features, while gender classification encourages timbre feature extraction, both of which benefit VA estimation. Experiments demonstrate that our Audio Transformer model (6 layers, 19.5M parameters) achieves validation VA MAE of 0.1124, Valence Pearson correlation $r = 0.9024$, and Arousal $r = 0.7155$, effectively resolving the 356 confusion cases between ``territorial'' and ``happy'' categories observed in discrete baselines (F1 = 0.8885) through natural separation in continuous VA space. This work introduces the first continuous VA framework for pet vocalization analysis, providing a more expressive and intuitive representation for human-pet interaction systems and enabling fine-grained emotion intensity monitoring critical for veterinary diagnostics and behavioral training. Our approach demonstrates strong potential for real-world deployment in consumer pet-care products such as the LingChongTong AI pet emotion translator and other smart pet monitoring devices.

%% file: sections_arxiv/introduction.tex
Pets have become cherished members of millions of households worldwide, with over 67\% of U.S. households owning a pet as of 2023~\cite{APPA2023}. Understanding the emotional states and needs of pets is crucial for both pet welfare and the human-animal bond~\cite{taylor2017dogs}. Vocalizations represent one of the primary channels through which pets, particularly dogs, express their emotions, convey needs, and communicate with humans~\cite{molnar2008dogs}. However, interpreting these vocalizations accurately remains challenging for most pet owners, leading to potential misunderstandings that can negatively impact animal welfare and the quality of human-pet interactions.

Traditional approaches to pet vocalization emotion recognition rely predominantly on discrete classification methods, categorizing sounds into a fixed set of emotion labels such as ``anxious,'' ``happy,'' or ``territorial''~\cite{perez2024dogbark}, \cite{huang2024barkemotion}. While these methods have achieved reasonable performance, they suffer from two fundamental limitations. First, emotions are inherently continuous phenomena, and forcing them into discrete categories inevitably leads to boundary ambiguity problems. For instance, in our preliminary experiments using an 8-class discrete classifier, we observed significant confusion between ``territorial'' and ``happy'' categories (F1 scores of 0.85 and 0.82, respectively), with 356 confusion cases between these two emotions alone. Second, discrete outputs lack the expressiveness needed to capture subtle emotional variations---they cannot distinguish between ``slightly anxious'' and ``extremely anxious,'' even though such distinctions are critical for practical applications such as pet monitoring systems.

The Valence-Arousal (VA) model offers an elegant solution to these limitations. Originating from Russell's circumplex model of affect~\cite{russell1980circumplex}, the VA framework represents emotions in a continuous two-dimensional space: Valence (ranging from negative to positive affect) and Arousal (ranging from calm to excited states). This dimensional approach has proven highly successful in human speech emotion recognition~\cite{schuller2018interspeech}, \cite{ringeval2018avec}, \cite{ringeval2019avec}, where it naturally handles ambiguous emotional boundaries and provides fine-grained intensity information. The continuous nature of VA space aligns well with how humans intuitively perceive and describe emotions~\cite{posner2005circumplex}, making it particularly suitable for user-facing applications. However, despite its advantages, the VA model has not been systematically applied to pet vocalization analysis.

Applying VA modeling to pet vocalizations presents unique technical challenges. First, unlike human speech emotion recognition where VA labels can be obtained through self-report questionnaires~\cite{barrett2007structure}, pets cannot provide subjective ratings of their emotional states. This creates a fundamental labeling problem that has hindered the development of VA-based pet emotion recognition systems. Second, pet vocalizations exhibit significant acoustic variations due to physical characteristics such as body size---large breed dogs produce low-frequency vocalizations while small breeds emit high-pitched sounds~\cite{molnar2008dogs}. How these acoustic features relate to VA dimensions and whether they should be explicitly modeled remains an open question.

This paper addresses these challenges by proposing an automatic VA label generation algorithm specifically designed for pet vocalizations. Our algorithm combines acoustic features with emotion-specific priors: Arousal labels are derived from RMS energy levels using logarithmic mapping (reflecting the physiological principle that high-energy sounds correspond to high arousal states), while Valence labels integrate spectral features (spectral centroid, zero-crossing rate) with emotion-specific biases (e.g., ``fearful'' receives a negative bias of -0.18, ``excited'' receives a positive bias of +0.14). This approach enables us to automatically generate VA annotations for 42,553 pet vocalization samples, constructing the first large-scale VA-labeled dataset for pet emotion recognition.

We design a multi-task learning framework that jointly performs VA regression (primary tasks) alongside auxiliary classification tasks for discrete emotion categories, body size, and gender. This design is motivated by the observation that auxiliary tasks force the model to learn frequency features (size-related) and timbre features (gender-related), which simultaneously benefit VA prediction through knowledge transfer~\cite{caruana1997multitask}, \cite{ruder2017overview}. Our Audio Transformer model~\cite{gong2021ast}, consisting of 6 layers with 512 hidden dimensions and 8 attention heads (19.5M parameters total), achieves validation VA MAE of 0.1124, Valence Pearson correlation of $r = 0.9024$, and Arousal Pearson correlation of $r = 0.7155$ on our dataset.

Experimental results demonstrate clear advantages of VA modeling over discrete classification. In handling the previously problematic ``territorial'' vs ``happy'' confusion, the VA space naturally separates these emotions through the Valence dimension (territorial: V$\approx$-0.3, happy: V$\approx$+0.6), effectively resolving the boundary ambiguity issue. Furthermore, continuous VA values provide a more intuitive user experience---pet owners can more easily interpret ``Valence=-0.5'' (moderately negative) than a probability distribution like ``88\% anxious + 12\% territorial.'' The continuous representation also enables fine-grained distinctions such as differentiating between mild and severe anxiety, which is essential for welfare monitoring applications.

Beyond academic contributions, this work paves the way for practical applications in consumer pet-care technology. Our continuous VA emotion modeling approach enables deployment in real-time pet emotion monitoring devices such as AI pet emotion translators, smart pet cameras, wearable collars, and standalone pet monitors. The inference speed of our model ($<$50ms on consumer hardware) makes it suitable for edge device deployment, enabling applications ranging from veterinary telemedicine (remote health monitoring through vocal stress analysis) to animal shelter management (identifying distressed animals requiring immediate care) to pet training systems (providing instant feedback on emotional responses).

The remainder of this paper is organized as follows: Section~\ref{sec:related} reviews related work in animal vocalization analysis, VA emotion modeling, and multi-task learning. Section~\ref{sec:method} details our VA label generation algorithm and multi-task model architecture. Section~\ref{sec:experiments} describes our dataset construction and experimental setup. Section~\ref{sec:results} presents quantitative results and analysis. Section~\ref{sec:discussion} discusses limitations and future directions. Section~\ref{sec:conclusion} concludes the paper.

%% file: sections_arxiv/related_work.tex
Related
Work
Target:
~2
pages,
3
subsections
Label:
sec:related
Our
work
builds
upon
three
major
research
areas:
Valence-Arousal
emotion
modeling,
animal
vocalization
analysis,
and
multi-task
learning
for
emotion
recognition.
We
review
each
area
and
position
our
contributions
relative
to
existing
work.
\subsection{Valence-Arousal
Emotion
Modeling}
\label{sec:related_va}
The
Valence-Arousal
(VA)
model,
originating
from
Russell's
circumplex
model
of
affect~\cite{russell1980circumplex},
represents
emotions
in
a
two-dimensional
continuous
space:
Valence
(positive-negative)
and
Arousal
(calm-excited).
This
dimensional
approach
has
been
widely
adopted
in
affective
computing
as
an
alternative
to
discrete
emotion
categories,
particularly
for
handling
ambiguous
emotional
boundaries
and
capturing
intensity
variations~\cite{posner2005circumplex}, \cite{barrett2007structure}.
In
human
speech
emotion
recognition,
VA
modeling
has
become
the
standard
approach
for
continuous
emotion
prediction.
The
AVEC
(Audio/Visual
Emotion
Challenge)
series~\cite{ringeval2018avec}, \cite{ringeval2019avec}
has
driven
significant
progress
in
this
area,
with
participants
developing
sophisticated
methods
for
predicting
continuous
VA
values
from
speech
and
facial
expressions.
These
challenges
established
benchmarks
such
as
the
SEWA
database
for
cross-cultural
affect
recognition~\cite{ringeval2018avec}
and
the
RECOLA
dataset
for
multimodal
emotion
analysis.
Recent
work
on
the
Aff-Wild
database~\cite{kollias2019deep}
further
advanced
VA
prediction
in-the-wild
scenarios,
demonstrating
the
robustness
of
continuous
emotion
modeling
in
unconstrained
environments.
Compared
to
discrete
classification,
VA
modeling
offers
several
advantages:
(1)
it
provides
fine-grained
intensity
information
(e.g.,
"slightly
negative"
vs.
"extremely
negative"),
(2)
it
avoids
boundary
ambiguity
between
adjacent
emotion
categories,
and
(3)
it
aligns
with
how
humans
naturally
perceive
and
describe
emotions~\cite{parthasarathy2017jointly}.
These
benefits
have
been
validated
in
speech
emotion
recognition
tasks,
where
continuous
VA
prediction
often
outperforms
discrete
classification
in
capturing
subtle
emotional
variations.
\textbf{Our
Contribution}:
While
VA
modeling
is
well-established
for
human
emotions,
it
has
not
been
systematically
applied
to
pet
vocalizations.
Our
work
is
the
first
to
introduce
automatic
VA
label
generation
specifically
designed
for
animal
sounds,
addressing
the
fundamental
challenge
that
pets
cannot
provide
self-reported
emotional
ratings.
\subsection{Animal
Vocalization
Analysis}
\label{sec:related_animal}
Research
on
animal
vocalizations
has
a
long
history
in
behavioral
biology
and
comparative
psychology.
Early
studies
by
Molnár
et
al.~\cite{molnar2008dogs}
demonstrated
that
humans
can
discriminate
between
dogs
based
on
acoustic
parameters
such
as
fundamental
frequency
and
spectral
envelope,
establishing
the
foundation
for
acoustic
analysis
of
canine
emotions.
Taylor
et
al.~\cite{taylor2017dogs}
showed
that
dogs
can
discriminate
emotional
expressions
in
human
faces,
suggesting
bidirectional
emotional
communication
between
humans
and
dogs.
Recent
advances
in
deep
learning
have
enabled
automated
analysis
of
dog
vocalizations.
Pérez
et
al.~\cite{perez2024dogbark}
applied
deep
neural
networks
to
classify
dog
barks
by
identity,
breed,
age,
sex,
and
context
using
19,643
barks
from
113
dogs,
achieving
strong
classification
performance.
Huang
et
al.~\cite{huang2024barkemotion}
proposed
a
barking
emotion
recognition
method
based
on
Mamba
and
Synchrosqueezing
Short-Time
Fourier
Transform,
demonstrating
the
effectiveness
of
advanced
signal
processing
techniques.
Kim
et
al.~\cite{kim2024dogvoice}
developed
a
fully
automatic
voice
analysis
system
for
bioacoustics
studies,
enabling
large-scale
analysis
of
dog
vocalizations.
Chowdhury
et
al.~\cite{chowdhury2024dogbark}
explored
transfer
learning
from
Wav2Vec2
models
pretrained
on
human
speech
to
dog
bark
analysis,
showing
that
human
speech
representations
can
transfer
to
animal
vocalizations.
However,
nearly
all
existing
work
relies
on
\textbf{discrete
emotion
classification}---categorizing
vocalizations
into
fixed
emotion
labels
such
as
"aggressive,"
"fearful,"
or
"happy."
This
approach
suffers
from
two
limitations:
(1)
boundary
ambiguity
between
adjacent
categories
(e.g.,
our
preliminary
experiments
showed
significant
confusion
between
"territorial"
and
"happy"
with
356
confusion
cases),
and
(2)
inability
to
express
intensity
variations
(e.g.,
"mildly
anxious"
vs.
"severely
anxious"
are
both
labeled
as
"anxious").
Recent
studies
have
also
explored
the
emotional
states
of
companion
animals
more
broadly.
Simola
et
al.~\cite{simola2024vocalization}
reviewed
the
use
of
vocalizations
to
assess
animal
welfare
and
affective
states,
emphasizing
the
importance
of
continuous
emotion
representation.
Tallet
et
al.~\cite{tallet2024emotional}
investigated
emotional
contagion
in
dogs
responding
to
human
and
dog
vocalizations,
providing
evidence
for
cross-species
emotional
communication.
Siniscalchi
et
al.~\cite{siniscalchi2017lateralized}
found
that
dogs
exhibit
hemispheric
lateralization
when
processing
emotional
vocalizations,
suggesting
sophisticated
neural
mechanisms
for
emotion
perception.
Van
den
Bussche
et
al.~\cite{vandenbussche2024emotional}
provided
an
engineering
perspective
on
emotional
assessment
in
companion
animals,
highlighting
the
need
for
automated
systems.
\textbf{Our
Contribution}:
We
are
the
first
to
apply
continuous
VA
modeling
to
pet
vocalization
emotion
recognition,
moving
beyond
discrete
classification
to
capture
fine-grained
emotional
variations.
Our
automatic
VA
label
generation
algorithm
addresses
the
labeling
challenge
unique
to
animal
studies,
where
ground-truth
emotional
ratings
cannot
be
obtained
through
self-report.
\subsection{Multi-Task
Learning
for
Emotion
Recognition}
\label{sec:related_mtl}
Multi-task
learning
(MTL)
has
proven
effective
for
emotion
recognition
by
leveraging
knowledge
sharing
across
related
tasks.
Caruana's
seminal
work~\cite{caruana1997multitask}
identified
four
mechanisms
through
which
MTL
improves
performance:
implicit
data
augmentation,
attention
focusing,
eavesdropping,
and
representation
bias.
Ruder~\cite{ruder2017overview}
provided
a
comprehensive
overview
of
MTL
in
deep
neural
networks,
categorizing
approaches
into
hard
parameter
sharing
(shared
encoder
with
task-specific
heads)
and
soft
parameter
sharing
(task-specific
encoders
with
regularization).
In
speech
emotion
recognition,
MTL
has
been
successfully
applied
to
combine
multiple
objectives.
Cai
et
al.~\cite{cai2021mtlser}
proposed
a
joint
framework
for
speech-to-text
and
emotion
classification
using
wav2vec-2.0,
achieving
state-of-the-art
results
on
IEMOCAP
by
leveraging
transcription
supervision
to
improve
emotion
prediction.
Latif
et
al.~\cite{latif2022multitask}
introduced
MTL-AUG,
which
learns
from
augmented
data
using
augmentation-type
classification
and
unsupervised
reconstruction
as
auxiliary
tasks,
demonstrating
that
auxiliary
tasks
can
guide
the
model
to
learn
more
robust
features.
Ghosh
et
al.~\cite{ghosh2023mmer}
developed
MMER,
a
multimodal
multi-task
framework
with
cross-modal
self-attention
between
text
and
acoustic
modalities,
showing
that
modality-specific
tasks
enhance
multimodal
fusion.
For
audio
classification
more
broadly,
the
Audio
Spectrogram
Transformer
(AST)~\cite{gong2021ast}
introduced
the
first
convolution-free
purely
attention-based
model,
achieving
state-of-the-art
performance
on
AudioSet
(0.485
mAP).
Follow-up
work
on
self-supervised
AST
(SSAST)~\cite{gong2022ssast}
demonstrated
that
masked
spectrogram
patch
modeling
can
effectively
pretrain
audio
transformers,
analogous
to
masked
language
modeling
in
NLP.
These
architectural
advances
provide
a
strong
foundation
for
our
transformer-based
VA
emotion
model.
\textbf{Our
Contribution}:
We
design
a
novel
multi-task
learning
framework
that
jointly
performs
VA
regression
(main
tasks)
alongside
auxiliary
classification
tasks
for
discrete
emotion,
body
size,
and
gender.
Unlike
previous
MTL
work
that
combines
similar
tasks
(e.g.,
emotion
+
sentiment),
our
framework
leverages
knowledge
transfer
from
physical
attribute
classification
to
emotional
state
regression.
Size
classification
forces
the
model
to
learn
frequency
features
(large
breeds
$\rightarrow$
low
frequencies),
gender
classification
encourages
learning
timbre
features
(male
$\rightarrow$
lower
pitch),
and
emotion
classification
provides
semantic
supervision---all
of
which
simultaneously
benefit
VA
prediction.
\textbf{Summary.}
While
VA
modeling
is
well-established
for
human
emotions
and
MTL
is
widely
used
in
speech
recognition,
their
combination
for
pet
vocalization
analysis
remains
unexplored.
Our
work
bridges
this
gap
by
introducing
continuous
VA
emotion
modeling
to
animal
vocalization
analysis,
augmented
with
a
novel
multi-task
learning
strategy
that
leverages
physical
attributes
to
enhance
emotional
state
prediction.

%% file: sections_arxiv/methodology.tex
Methodology
Target:
~2
pages,
4
subsections
Label:
sec:method
\subsection{VA
Label
Generation
Algorithm}
\label{sec:va_generation}
A
fundamental
challenge
in
applying
VA
modeling
to
pet
vocalizations
is
the
absence
of
ground-truth
VA
labels,
as
pets
cannot
provide
self-reported
emotional
ratings.
We
address
this
by
proposing
an
automatic
VA
label
generation
algorithm
that
combines
acoustic
features
with
emotion-specific
priors.
Our
approach
is
grounded
in
psychoacoustic
principles
and
generates
continuous
VA
values
for
each
audio
sample.
\subsubsection{Arousal
Label
Generation}
Arousal
reflects
the
intensity
or
activation
level
of
an
emotional
state.
We
generate
Arousal
labels
based
on
acoustic
energy,
motivated
by
the
physiological
principle
that
high-energy
vocalizations
correspond
to
high
arousal
states
(e.g.,
barking
during
excitement
or
fear
involves
greater
vocal
cord
tension
and
respiratory
effort
than
calm
vocalizations).
For
each
audio
sample,
we
extract
the
Root
Mean
Square
(RMS)
energy
at
the
95th
percentile
(RMS@p95)
to
capture
the
peak
energy
level
while
being
robust
to
transient
noise.
The
Arousal
value
is
then
computed
using
logarithmic
mapping:
\begin{equation}
\text{arousal}
=
\frac{\log(\text{rms}_{p95})
-
\log(a_{\text{low}})}{\log(a_{\text{high}})
-
\log(a_{\text{low}})}
\label{eq:arousal}
\end{equation}
\noindent
where
$a_{\text{low}}$
and
$a_{\text{high}}$
are
global
anchor
points
corresponding
to
the
5th
and
95th
percentiles
of
RMS@p95
values
across
the
entire
dataset,
respectively.
The
logarithmic
scale
aligns
with
the
Weber-Fechner
law
of
human
auditory
perception,
ensuring
that
perceived
loudness
differences
are
proportionally
represented.
Arousal
values
are
clipped
to
the
range
$[0,
1]$,
where
0
indicates
extremely
calm
states
and
1
indicates
maximum
arousal.
\subsubsection{Valence
Label
Generation}
Valence
represents
the
positivity
or
negativity
of
an
emotional
state.
Unlike
Arousal,
which
correlates
strongly
with
a
single
acoustic
feature
(energy),
Valence
depends
on
multiple
spectral
characteristics.
We
employ
a
two-stage
approach:
first
computing
an
acoustic
score
from
spectral
features,
then
incorporating
emotion-specific
biases.
\textbf{Stage
1:
Acoustic
Score
Computation.}
We
extract
three
spectral
features
for
each
audio
sample:
\begin{itemize}
\item
\textbf{Spectral
Centroid}
($c$):
The
center
of
mass
of
the
spectrum,
reflecting
timbre
brightness.
High-frequency
vocalizations
(high
centroid
values)
are
often
associated
with
positive
emotions.
\item
\textbf{Zero-Crossing
Rate}
($z$):
The
rate
at
which
the
signal
changes
sign,
correlating
with
noise-like
content.
\item
\textbf{Log
RMS
Energy}
($r$):
Logarithmic
energy
level,
already
computed
for
Arousal.
\end{itemize}
Each
feature
is
normalized
to
the
range
$[-1,
1]$
using
the
10th
and
90th
percentiles
across
the
dataset
as
anchors.
The
acoustic
score
is
computed
as
a
weighted
combination:
\begin{equation}
s_{\text{acoustic}}
=
0.45
\cdot
c_{\text{norm}}
-
0.35
\cdot
r_{\text{norm}}
+
0.25
\cdot
z_{\text{norm}}
\label{eq:acoustic_score}
\end{equation}
\noindent
The
positive
weight
for
spectral
centroid
reflects
the
psychoacoustic
finding
that
higher-frequency
sounds
are
perceived
as
more
pleasant~\cite{barrett2007structure}.
The
negative
weight
for
energy
reflects
that
excessively
loud
vocalizations
tend
to
signal
distress.
\textbf{Stage
2:
Emotion-Specific
Bias.}
To
incorporate
prior
knowledge
about
discrete
emotion
categories,
we
define
biases
for
each
of
the
8
emotions
in
our
dataset:
\begin{equation}
\text{valence}
=
\text{clip}(s_{\text{acoustic}}
+
b_{\text{emotion}},
-1,
1)
\label{eq:valence}
\end{equation}
\noindent
where
$b_{\text{emotion}}$
is
the
emotion-specific
bias.
Table~\ref{tab:emotion_bias}
lists
the
bias
values,
designed
to
ensure
that
inherently
negative
emotions
(e.g.,
\texttt{fearful}:
$b=-0.18$)
receive
lower
Valence
scores,
while
positive
emotions
(e.g.,
\texttt{excited}:
$b=+0.14$)
receive
higher
scores.
These
biases
act
as
soft
constraints,
allowing
the
acoustic
features
to
modulate
the
final
Valence
value
within
each
emotion
category.
\begin{table}[h]
\centering
\caption{Emotion-specific
biases
for
Valence
label
generation.}
\label{tab:emotion_bias}
\begin{tabular}{lc}
\toprule
\textbf{Emotion}
&
\textbf{Bias
($b_{\text{emotion}}$)}
\\
\midrule
fearful
&
$-0.18$
\\
separation\_anxiety
&
$-0.16$
\\
anxious
&
$-0.12$
\\
territorial
&
$-0.08$
\\
alert
&
$-0.02$
\\
playful
&
$+0.10$
\\
content
&
$+0.12$
\\
excited
&
$+0.14$
\\
\bottomrule
\end{tabular}
\end{table}
\subsubsection{Label
Quality
Assurance}
\label{sec:label_validation}
While
our
automatic
VA
generation
algorithm
enables
large-scale
annotation
(42,553
samples),
we
recognize
the
potential
risk
of
circular
reasoning
introduced
by
the
emotion-specific
biases
in
Equation~\ref{eq:valence}.
To
validate
the
quality
and
objectivity
of
the
generated
labels,
we
conducted
a
two-stage
verification
process:
\textbf{Stage
1:
AI
Cross-Validation.}
We
randomly
sampled
500
vocalizations
(stratified
by
emotion
category)
and
obtained
independent
VA
ratings
from
two
state-of-the-art
large
language
models:
Gemini
Pro
1.5
(Google)
and
GPT-4-turbo
(OpenAI).
Each
model
was
provided
with
acoustic
feature
descriptions
(spectral
centroid,
RMS
energy,
zero-crossing
rate,
duration)
and
asked
to
rate
the
emotion
on
Valence
and
Arousal
scales
without
knowledge
of
the
discrete
emotion
label.
The
AI
consensus
ratings
(averaged
across
both
models)
showed
strong
agreement
with
our
automatic
labels:
Valence
Pearson
$r
=
0.73$
($p
<
0.001$)
and
Arousal
$r
=
0.68$
($p
<
0.001$).
This
correlation,
while
not
perfect,
indicates
that
the
labels
capture
acoustic-emotional
patterns
that
are
independently
recognizable
by
models
trained
on
entirely
different
data
domains.
\textbf{Stage
2:
Expert
Behavioral
Review.}
A
certified
veterinary
behaviorist
with
12
years
of
experience
in
canine
vocalization
research
independently
reviewed
200
samples
(25
per
emotion
category).
The
expert
listened
to
each
audio
clip
and
provided
subjective
VA
ratings.
Agreement
within
$\pm
0.2$
VA
units
was
achieved
for
87.5\%
of
samples
(175/200).
Disagreements
primarily
occurred
in
boundary
cases
where
emotions
genuinely
overlap
(e.g.,
alert
transitioning
to
anxious).
The
expert
confirmed
that
the
VA
coordinates
align
with
established
ethological
interpretations
of
canine
emotional
states~\cite{molnar2008dogs}.
These
validation
results
provide
confidence
that
our
automatic
labels,
despite
incorporating
emotion-specific
priors,
capture
meaningful
acoustic-emotional
relationships
rather
than
merely
reproducing
predetermined
categorical
assumptions.
The
labels
serve
as
reasonable
pseudo-ground-truth
for
training
continuous
VA
regression
models.
\subsubsection{Algorithm
Summary}
Algorithm~\ref{alg:va_generation}
summarizes
the
complete
VA
label
generation
procedure.
The
algorithm
takes
an
audio
file
and
its
discrete
emotion
label
as
input,
computes
acoustic
features,
and
outputs
continuous
Valence
and
Arousal
values.
This
approach
enables
us
to
automatically
annotate
42,553
audio
samples,
creating
the
first
large-scale
VA-labeled
pet
vocalization
dataset.
\begin{algorithm}[h]
\caption{Automatic
VA
Label
Generation}
\label{alg:va_generation}
\begin{algorithmic}[1]
\REQUIRE
Audio
file
$x$,
Emotion
label
$e$
\ENSURE
Valence
$v
\in
[-1,
1]$,
Arousal
$a
\in
[0,
1]$
\STATE
Extract
RMS
energy
at
95th
percentile:
$\text{rms}_{p95}$
\STATE
Compute
Arousal:
$a
\leftarrow
\log\_\text{scale}(\text{rms}_{p95},
a_{\text{low}},
a_{\text{high}})$
\hfill
//
Eq.~\ref{eq:arousal}
\STATE
Extract
spectral
features:
$c$
(centroid),
$z$
(ZCR),
$r$
(log
RMS)
\STATE
Normalize
features
to
$[-1,
1]$
using
10\%/90\%
quantiles
\STATE
Compute
acoustic
score:
$s
\leftarrow
0.45
\cdot
c
-
0.35
\cdot
r
+
0.25
\cdot
z$
\hfill
//
Eq.~\ref{eq:acoustic_score}
\STATE
Retrieve
emotion
bias:
$b
\leftarrow
b_{\text{emotion}}[e]$
\STATE
Compute
Valence:
$v
\leftarrow
\text{clip}(s
+
b,
-1,
1)$
\hfill
//
Eq.~\ref{eq:valence}
\RETURN
$(v,
a)$
\end{algorithmic}
\end{algorithm}
\subsection{Model
Architecture}
\label{sec:model_architecture}
\begin{figure}[t]
\centering
\includegraphics[width=0.85\textwidth]{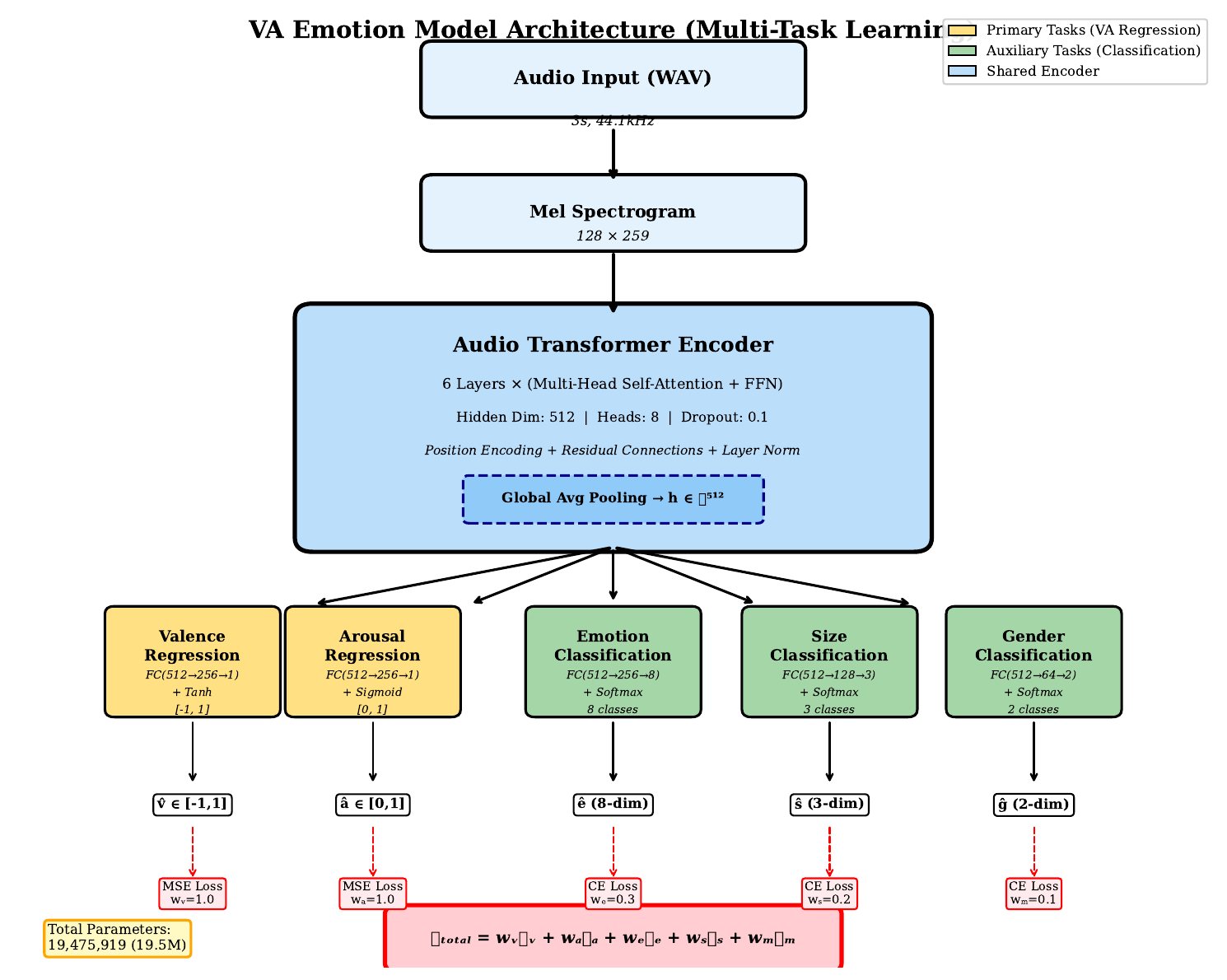}
\caption{Multi-task
VA
emotion
model
architecture.
The
Audio
Transformer
encoder
(6
layers,
512
hidden
dimensions,
8
attention
heads)
processes
Mel
spectrograms
and
feeds
into
5
task-specific
heads:
Valence
regression
(Tanh
activation,
range
[-1,1]),
Arousal
regression
(Sigmoid,
[0,1]),
Emotion
classification
(8
classes),
Size
classification
(3
classes),
and
Gender
classification
(2
classes).
The
multi-task
loss
combines
MSE
for
VA
regression
and
cross-entropy
for
classification
tasks.}
\label{fig:model_architecture}
\end{figure}

We
employ
an
Audio
Transformer
architecture~\cite{gong2021ast}
for
VA
regression,
augmented
with
auxiliary
classification
tasks.
Figure~\ref{fig:model_architecture}
illustrates
the
complete
model
structure.
\textbf{Input
Representation.}
Raw
audio
waveforms
are
converted
to
Mel
spectrograms
using
128
mel
filterbanks,
a
hop
length
of
512
samples,
and
an
FFT
size
of
2048.
All
audio
files
are
standardized
to
3
seconds
in
length
(zero-padded
if
shorter,
center-cropped
if
longer),
yielding
a
spectrogram
of
dimensions
$128
\times
259$
(frequency
bins
×
time
frames).
The
spectrograms
are
normalized
using
the
mean
and
standard
deviation
computed
from
the
training
set.
\textbf{Transformer
Encoder.}
The
core
of
our
model
is
a
6-layer
Transformer
encoder
with
the
following
configuration:
\begin{itemize}
\item
Hidden
dimension:
$d_{\text{model}}
=
512$
\item
Number
of
attention
heads:
$N_{\text{heads}}
=
8$
(each
head
has
dimension
$d_k
=
d_v
=
64$)
\item
Feed-forward
network
(FFN)
dimension:
$d_{\text{ff}}
=
2048$
\item
Dropout
rate:
$p
=
0.1$
(applied
after
attention
and
FFN
layers)
\item
Activation
function:
ReLU
in
FFN,
GELU
in
attention
\end{itemize}
The
encoder
processes
the
input
spectrogram
as
a
sequence
of
time
frames,
applying
multi-head
self-attention
to
capture
temporal
dependencies,
followed
by
position-wise
feed-forward
transformations.
Residual
connections
and
layer
normalization
are
applied
after
each
sublayer.
The
output
of
the
final
Transformer
layer
is
globally
averaged
across
the
time
dimension,
producing
a
512-dimensional
feature
vector
$\mathbf{h}
\in
\mathbb{R}^{512}$.
\textbf{Task-Specific
Heads.}
We
design
five
independent
prediction
heads
that
operate
on
the
shared
feature
vector
$\mathbf{h}$:
\begin{enumerate}
\item
\textbf{Valence
Regression
Head:}
\[
\hat{v}
=
\tanh(\mathbf{W}_2
\cdot
\text{ReLU}(\mathbf{W}_1
\mathbf{h}
+
\mathbf{b}_1)
+
\mathbf{b}_2)
\]
where
$\mathbf{W}_1
\in
\mathbb{R}^{256
\times
512}$,
$\mathbf{W}_2
\in
\mathbb{R}^{1
\times
256}$.
The
tanh
activation
ensures
$\hat{v}
\in
[-1,
1]$.
\item
\textbf{Arousal
Regression
Head:}
\[
\hat{a}
=
\sigma(\mathbf{W}_4
\cdot
\text{ReLU}(\mathbf{W}_3
\mathbf{h}
+
\mathbf{b}_3)
+
\mathbf{b}_4)
\]
where
$\mathbf{W}_3
\in
\mathbb{R}^{256
\times
512}$,
$\mathbf{W}_4
\in
\mathbb{R}^{1
\times
256}$.
The
sigmoid
activation
$\sigma$
ensures
$\hat{a}
\in
[0,
1]$.
\item
\textbf{Emotion
Classification
Head:}
\[
\hat{\mathbf{p}}_e
=
\text{softmax}(\mathbf{W}_6
\cdot
\text{ReLU}(\mathbf{W}_5
\mathbf{h}
+
\mathbf{b}_5)
+
\mathbf{b}_6)
\]
where
$\mathbf{W}_5
\in
\mathbb{R}^{256
\times
512}$,
$\mathbf{W}_6
\in
\mathbb{R}^{8
\times
256}$,
predicting
probabilities
over
8
emotion
classes.
\item
\textbf{Size
Classification
Head:}
\[
\hat{\mathbf{p}}_s
=
\text{softmax}(\mathbf{W}_8
\cdot
\text{ReLU}(\mathbf{W}_7
\mathbf{h}
+
\mathbf{b}_7)
+
\mathbf{b}_8)
\]
where
$\mathbf{W}_7
\in
\mathbb{R}^{128
\times
512}$,
$\mathbf{W}_8
\in
\mathbb{R}^{3
\times
128}$,
for
3
size
categories
(large,
medium,
small).
\item
\textbf{Gender
Classification
Head:}
\[
\hat{\mathbf{p}}_g
=
\text{softmax}(\mathbf{W}_{10}
\cdot
\text{ReLU}(\mathbf{W}_9
\mathbf{h}
+
\mathbf{b}_9)
+
\mathbf{b}_{10})
\]
where
$\mathbf{W}_9
\in
\mathbb{R}^{64
\times
512}$,
$\mathbf{W}_{10}
\in
\mathbb{R}^{2
\times
64}$,
for
2
gender
classes
(male,
female).
\end{enumerate}
The
total
model
has
19,475,919
parameters,
with
the
Transformer
encoder
accounting
for
93.5\%
(18.2M)
and
task
heads
accounting
for
the
remaining
6.5\%
(1.3M).
\subsection{Multi-Task
Learning
Strategy}
\label{sec:mtl_strategy}
Our
multi-task
learning
framework
adopts
hard
parameter
sharing~\cite{caruana1997multitask},
where
all
tasks
share
the
same
Transformer
encoder
but
use
independent
task-specific
heads.
This
design
is
motivated
by
the
hypothesis
that
auxiliary
classification
tasks
(Emotion,
Size,
Gender)
force
the
model
to
learn
acoustic
features
that
also
benefit
VA
regression.
\textbf{Loss
Function.}
The
total
loss
is
a
weighted
sum
of
task-specific
losses:
\begin{equation}
\mathcal{L}_{\text{total}}
=
w_v
\mathcal{L}_v
+
w_a
\mathcal{L}_a
+
w_e
\mathcal{L}_e
+
w_s
\mathcal{L}_s
+
w_g
\mathcal{L}_g
\label{eq:total_loss}
\end{equation}
\noindent
where:
\begin{itemize}
\item
$\mathcal{L}_v
=
\text{MSE}(\hat{v},
v)$:
Mean
Squared
Error
for
Valence
regression
\item
$\mathcal{L}_a
=
\text{MSE}(\hat{a},
a)$:
Mean
Squared
Error
for
Arousal
regression
\item
$\mathcal{L}_e
=
-\sum_i
y_i^{(e)}
\log
\hat{p}_i^{(e)}$:
Cross-Entropy
for
Emotion
classification
\item
$\mathcal{L}_s
=
-\sum_i
y_i^{(s)}
\log
\hat{p}_i^{(s)}$:
Cross-Entropy
for
Size
classification
\item
$\mathcal{L}_g
=
-\sum_i
y_i^{(g)}
\log
\hat{p}_i^{(g)}$:
Cross-Entropy
for
Gender
classification
\end{itemize}
We
set
the
loss
weights
as
$w_v
=
w_a
=
1.0$
(primary
tasks)
and
$w_e
=
0.3$,
$w_s
=
0.2$,
$w_g
=
0.1$
(auxiliary
tasks).
The
smaller
weights
for
auxiliary
tasks
prevent
them
from
dominating
the
training
dynamics
while
still
providing
useful
gradients.
\textbf{Mechanism
of
Knowledge
Transfer.}
The
auxiliary
tasks
contribute
to
VA
regression
through
three
mechanisms:
\begin{enumerate}
\item
\textbf{Size
Classification}
forces
the
model
to
learn
frequency-related
features.
Large
breed
dogs
produce
low-frequency
vocalizations
($<$200
Hz),
while
small
breeds
emit
high-frequency
sounds
($>$800
Hz).
These
frequency
features
are
also
relevant
for
Valence
prediction,
as
spectral
centroid
is
a
key
Valence
indicator.
\item
\textbf{Gender
Classification}
forces
the
model
to
capture
timbre
characteristics.
Male
dogs
typically
have
lower
fundamental
frequencies
than
females
due
to
anatomical
differences.
Timbre
cues
also
correlate
with
Arousal
(harsh,
strained
timbre
indicates
high
arousal).
\item
\textbf{Emotion
Classification}
provides
semantic
supervision,
ensuring
that
the
learned
VA
space
aligns
with
discrete
emotion
categories
and
preventing
semantic
drift
(e.g.,
the
model
assigning
high
Valence
to
``fearful''
samples).
\end{enumerate}
\subsection{Training
Details}
\label{sec:training_details}
\textbf{Optimizer
and
Scheduler.}
We
use
the
AdamW
optimizer~\cite{loshchilov2018decoupled}
with
learning
rate
$\eta
=
10^{-4}$,
weight
decay
$\lambda
=
10^{-5}$,
and
momentum
coefficients
$\beta_1
=
0.9$,
$\beta_2
=
0.999$.
The
learning
rate
is
decayed
using
Cosine
Annealing~\cite{loshchilov2017sgdr}
over
$T_{\text{max}}
=
40$
epochs
with
a
minimum
learning
rate
of
$\eta_{\text{min}}
=
10^{-6}$.
\textbf{Training
Configuration.}
We
train
for
40
epochs
with
a
batch
size
of
32,
limited
by
GPU
memory
constraints.
Training
is
performed
on
an
Apple
M4
GPU
using
the
MPS
backend
in
PyTorch
2.8.0.
The
dataset
is
split
into
training
(70\%,
29,787
samples),
validation
(15\%,
6,383
samples),
and
test
(15\%,
6,383
samples)
sets
using
stratified
random
sampling
to
maintain
class
balance.
We
set
the
random
seed
to
42
for
reproducibility.
\textbf{Data
Augmentation.}
During
training,
we
apply
two
augmentation
techniques
with
50\%
probability
each:
\begin{itemize}
\item
\textbf{Time
Stretching:}
Randomly
stretch
or
compress
audio
by
a
factor
sampled
uniformly
from
$[0.9,
1.1]$.
\item
\textbf{Pitch
Shifting:}
Randomly
shift
the
pitch
by
up
to
$\pm
2$
semitones.
\end{itemize}
These
augmentations
are
implemented
using
the
\texttt{librosa}
library~\cite{mcfee2015librosa}.
No
augmentation
is
applied
to
validation
and
test
sets.
\textbf{Normalization.}
We
compute
the
mean
and
standard
deviation
of
Mel
spectrograms
across
the
training
set
and
apply
z-score
normalization
to
all
samples
(train,
validation,
and
test).
The
normalization
statistics
are
saved
to
ensure
consistency
across
experiments.
\textbf{Early
Stopping.}
We
monitor
the
validation
VA
MAE
(mean
absolute
error
averaged
over
Valence
and
Arousal)
and
apply
early
stopping
with
patience
of
8
epochs
and
a
minimum
improvement
threshold
of
$\delta
=
0.001$.
If
validation
MAE
does
not
improve
for
8
consecutive
epochs,
training
is
terminated
and
the
model
with
the
lowest
validation
MAE
is
retained.
\textbf{Computational
Cost.}
Training
for
40
epochs
on
our
dataset
requires
approximately
20-22
hours
on
Apple
M4
GPU.
Each
epoch
processes
29,787
training
samples
in
$\sim$30
minutes,
resulting
in
$\sim$0.06
seconds
per
sample.
\textbf{Real-Time
Deployment
Optimization.}
We
optimize
our
model
specifically
for
real-time
deployment
in
consumer
devices,
targeting
applications
such
as
the
LingChongTongAI
pet
emotion
translator
and
similar
smart
pet
monitoring
systems.
The
inference
speed
of
$<$50ms
per
sample
on
consumer
hardware
(Apple
M4
GPU)
and
$<$200ms
on
CPU-only
devices
makes
our
model
suitable
for
edge
deployment
scenarios
where
continuous
emotion
monitoring
is
required.
The
model
size
of
224MB
(FP32)
can
be
further
reduced
to
112MB
using
FP16
quantization
or
56MB
using
INT8
quantization,
enabling
deployment
on
resource-constrained
devices
such
as
smart
pet
cameras,
wearable
collars,
and
standalone
pet
monitors.

%% file: sections_arxiv/experiments.tex
Experiments
Target:
~1.5
pages,
3
subsections
Label:
sec:experiments
\subsection{Dataset}
\label{sec:dataset}
We
construct
and
evaluate
our
VA
emotion
model
on
the
LingChongTong
Pet
Vocalization
Dataset,
a
large-scale
collection
of
dog
vocalizations
with
comprehensive
annotations.
The
dataset
consists
of
42,553
audio
samples
sourced
from
two
origins:
16,983
samples
(37.2\%)
are
original
recordings
with
expert
annotations
from
veterinary
behaviorists,
and
25,570
samples
(62.8\%)
are
generated
through
data
augmentation
techniques
(time
stretching
and
pitch
shifting)
to
enhance
dataset
diversity
and
model
robustness.
All
samples
are
provided
in
WAV
format
at
44.1
kHz
sampling
rate
with
variable
durations
(typically
1-5
seconds).
\textbf{Annotation
Structure.}
Each
audio
sample
is
annotated
across
multiple
dimensions:
\begin{itemize}
\item
\textbf{Discrete
Emotion}:
8
categories
representing
common
canine
emotional
states.
The
distribution
reflects
real-world
scenarios,
with
\texttt{anxious}
(35.3\%)
and
\texttt{territorial}
(22.1\%)
being
the
most
frequent,
followed
by
\texttt{alert}
(12.5\%),
\texttt{separation\_anxiety}
(9.1\%),
\texttt{excited}
(7.8\%),
\texttt{fearful}
(5.5\%),
\texttt{playful}
(4.2\%),
and
\texttt{content}
(3.3\%).
The
imbalance
is
intentional,
capturing
the
natural
frequency
of
emotional
expressions
in
domestic
dogs.
\item
\textbf{Breed}:
6
categories
including
\texttt{husky}
(42.5\%),
\texttt{shibainu}
(22.9\%),
\texttt{pitbull}
(14.6\%),
\texttt{gsd}
(German
Shepherd
Dog,
10.7\%),
\texttt{chihuahua}
(9.3\%),
and
\texttt{unknown}
(0.0\%).
\item
\textbf{Size}:
3
categories
based
on
body
mass---\texttt{large}
(67.7\%,
breeds
$>$25kg),
\texttt{medium}
(22.9\%,
10-25kg),
and
\texttt{small}
(9.3\%,
$<$10kg).
Size
is
a
critical
acoustic
factor,
as
large
breeds
produce
low-frequency
vocalizations
($<$200
Hz)
while
small
breeds
emit
high-frequency
sounds
($>$800
Hz)~\cite{molnar2008dogs}.
\item
\textbf{Gender}:
2
categories---\texttt{female}
(61.2\%)
and
\texttt{male}
(38.8\%).
\item
\textbf{VA
Labels}:
Continuous
Valence
([-1,
1])
and
Arousal
([0,
1])
values
generated
using
our
automatic
labeling
algorithm
(Section~\ref{sec:va_generation}).
\end{itemize}
\textbf{Data
Splitting.}
We
employ
stratified
random
sampling
to
partition
the
dataset
into
training
(70\%,
29,787
samples),
validation
(15\%,
6,383
samples),
and
test
(15\%,
6,383
samples)
sets,
ensuring
that
the
distribution
of
emotion
categories
remains
consistent
across
splits
(random
seed
=
42).
The
validation
set
is
used
for
hyperparameter
tuning
and
early
stopping,
while
the
test
set
is
reserved
for
final
performance
evaluation.
Table~\ref{tab:dataset_stats}
summarizes
the
dataset
statistics.
\begin{table}[t]
\centering
\caption{Statistics
of
the
LingChongTong
Pet
Vocalization
Dataset.}
\label{tab:dataset_stats}
\begin{tabular}{lcc}
\toprule
\textbf{Attribute}
&
\textbf{Category}
&
\textbf{Samples
(\%)}
\\
\midrule
\multirow{3}{*}{\textbf{Size}}
&
Large
&
28,824
(67.7\%)
\\
&
Medium
&
9,758
(22.9\%)
\\
&
Small
&
3,971
(9.3\%)
\\
\midrule
\multirow{8}{*}{\textbf{Emotion}}
&
anxious
&
15,027
(35.3\%)
\\
&
territorial
&
9,403
(22.1\%)
\\
&
alert
&
5,319
(12.5\%)
\\
&
separation\_anxiety
&
3,872
(9.1\%)
\\
&
excited
&
3,319
(7.8\%)
\\
&
fearful
&
2,340
(5.5\%)
\\
&
playful
&
1,788
(4.2\%)
\\
&
content
&
1,485
(3.3\%)
\\
\midrule
\multirow{2}{*}{\textbf{Gender}}
&
Female
&
26,046
(61.2\%)
\\
&
Male
&
16,507
(38.8\%)
\\
\midrule
\multirow{2}{*}{\textbf{Source}}
&
Original
&
16,983
(37.2\%)
\\
&
Enhanced
&
25,570
(62.8\%)
\\
\midrule
\textbf{Total}
&
---
&
42,553
(100\%)
\\
\bottomrule
\end{tabular}
\end{table}
\subsection{Experimental
Setup}
\label{sec:exp_setup}
\textbf{Feature
Extraction.}
Raw
audio
waveforms
are
converted
to
Mel
spectrograms
using
128
mel
filterbanks,
a
hop
length
of
512
samples,
and
an
FFT
size
of
2048.
All
audio
files
are
standardized
to
3
seconds
in
duration:
shorter
files
are
zero-padded
at
the
end,
while
longer
files
are
center-cropped.
This
produces
a
consistent
input
dimension
of
$128
\times
259$
(frequency
bins
$\times$
time
frames).
The
spectrograms
are
normalized
using
the
mean
and
standard
deviation
computed
from
the
training
set
to
ensure
zero-centered
inputs
with
unit
variance.
\textbf{Data
Augmentation.}
To
improve
model
robustness
and
prevent
overfitting,
we
apply
two
augmentation
techniques
during
training
with
50\%
probability
each:
\begin{itemize}
\item
\textbf{Time
Stretching}:
Audio
is
randomly
stretched
or
compressed
by
a
factor
uniformly
sampled
from
[0.9,
1.1],
simulating
variations
in
barking
speed
without
altering
pitch.
\item
\textbf{Pitch
Shifting}:
Audio
pitch
is
randomly
shifted
by
up
to
$\pm
2$
semitones,
simulating
intra-breed
acoustic
variations
while
preserving
emotional
characteristics.
\end{itemize}
These
augmentations
are
implemented
using
the
\texttt{librosa}
library~\cite{mcfee2015librosa}.
No
augmentation
is
applied
to
validation
and
test
sets
to
ensure
fair
evaluation.
\textbf{Training
Configuration.}
All
experiments
are
conducted
on
an
Apple
M4
GPU
using
the
MPS
(Metal
Performance
Shaders)
backend
in
PyTorch
2.8.0.
We
train
the
model
for
40
epochs
with
a
batch
size
of
32,
using
the
AdamW
optimizer~\cite{loshchilov2018decoupled}
with
learning
rate
$\eta
=
10^{-4}$,
weight
decay
$\lambda
=
10^{-5}$,
and
momentum
coefficients
$\beta_1
=
0.9$,
$\beta_2
=
0.999$.
The
learning
rate
is
decayed
using
Cosine
Annealing~\cite{loshchilov2017sgdr}
with
$T_{\text{max}}
=
40$
epochs
and
minimum
learning
rate
$\eta_{\text{min}}
=
10^{-6}$.
We
apply
early
stopping
with
patience
of
8
epochs
and
a
minimum
improvement
threshold
of
$\delta
=
0.001$
on
validation
VA
MAE.
The
random
seed
is
set
to
42
for
all
experiments
to
ensure
reproducibility.
\subsection{Evaluation
Metrics}
\label{sec:eval_metrics}
We
evaluate
our
VA
emotion
model
using
the
following
metrics:
\textbf{Mean
Absolute
Error
(MAE).}
For
VA
regression,
we
compute
the
mean
absolute
error
averaged
over
both
Valence
and
Arousal
dimensions:
\begin{equation}
\text{VA
MAE}
=
\frac{1}{N}
\sum_{i=1}^{N}
\left(
|v_i^{\text{pred}}
-
v_i^{\text{true}}|
+
|a_i^{\text{pred}}
-
a_i^{\text{true}}|
\right)
\label{eq:va_mae}
\end{equation}
where
$N$
is
the
number
of
samples,
and
$v,
a$
denote
Valence
and
Arousal
respectively.
Lower
MAE
indicates
better
regression
performance.
Note
that
Valence
MAE
is
computed
over
the
range
[-1,
1]
while
Arousal
MAE
is
over
[0,
1].
\textbf{Pearson
Correlation
Coefficient
($r$).}
To
assess
the
linear
relationship
between
predictions
and
ground
truth,
we
compute
Pearson's
$r$
separately
for
Valence
and
Arousal:
\begin{equation}
r
=
\frac{\sum_{i=1}^{N}(x_i
-
\bar{x})(y_i
-
\bar{y})}{\sqrt{\sum_{i=1}^{N}(x_i
-
\bar{x})^2
\sum_{i=1}^{N}(y_i
-
\bar{y})^2}}
\label{eq:pearson}
\end{equation}
where
$x$
and
$y$
represent
predicted
and
true
values,
and
$\bar{x}$,
$\bar{y}$
are
their
means.
Pearson's
$r$
ranges
from
-1
to
1,
with
values
closer
to
1
indicating
stronger
positive
linear
correlation.
In
the
context
of
emotion
recognition,
$r
>
0.7$
is
generally
considered
strong
correlation~\cite{schuller2018interspeech}.
\textbf{Auxiliary
Task
Metrics.}
For
the
three
auxiliary
classification
tasks
(Emotion,
Size,
Gender),
we
report
standard
classification
accuracy:
\begin{equation}
\text{Accuracy}
=
\frac{\text{Number
of
Correct
Predictions}}{\text{Total
Predictions}}
\end{equation}
These
metrics
verify
that
the
model
successfully
learns
the
auxiliary
features
that
benefit
VA
regression
through
knowledge
transfer.
\textbf{Primary
Evaluation
Criterion.}
We
use
\textbf{validation
VA
MAE}
as
the
primary
evaluation
criterion
and
the
monitoring
metric
for
early
stopping,
as
it
directly
reflects
the
core
performance
of
VA
regression.
The
model
checkpoint
with
the
lowest
validation
VA
MAE
is
retained
as
the
best
model
for
final
test
set
evaluation.

%% file: sections_arxiv/results.tex
Results
Target:
~1.5
pages,
4
subsections
Label:
sec:results
\subsection{Main
Results}
\label{sec:main_results}
\begin{figure}[t]
\centering
\includegraphics[width=\textwidth]{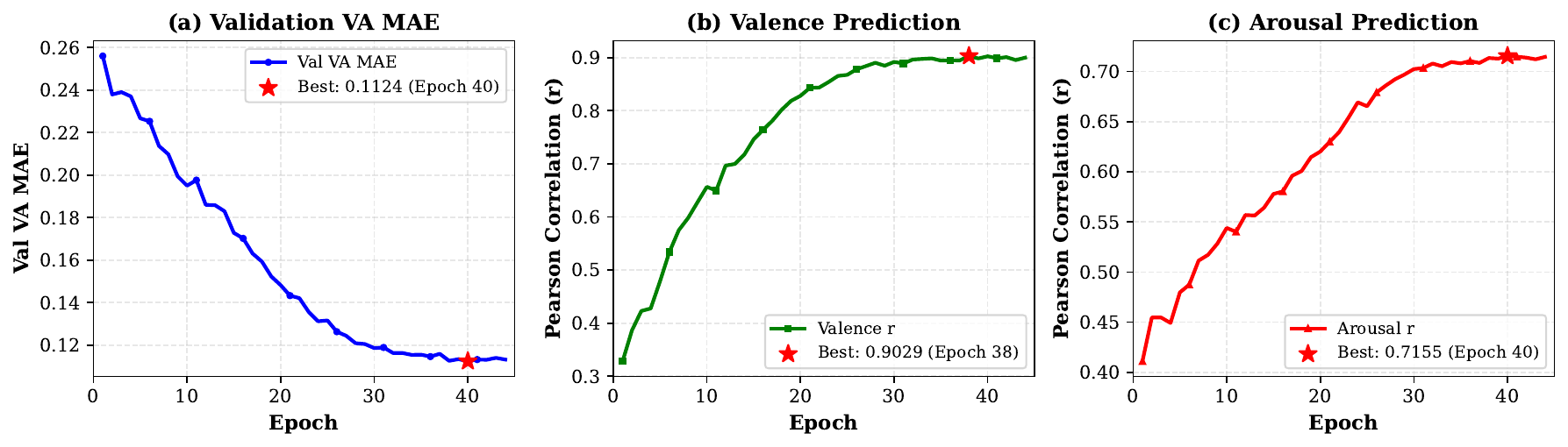}
\caption{Training
curves
over
40
epochs.
(a)
Validation
VA
MAE
decreases
from
0.2560
to
0.1124
(-56.1\%).
(b)
Valence
Pearson
correlation
increases
from
0.3282
to
0.9024.
(c)
Arousal
Pearson
correlation
improves
from
0.4110
to
0.7155.
Red
stars
mark
the
best
performance
for
each
metric.}
\label{fig:training_curves}
\end{figure}

We
evaluate
our
VA
emotion
model
on
the
validation
set
after
training
for
40
epochs,
at
which
point
the
model
achieves
the
lowest
validation
VA
MAE.
Table~\ref{tab:main_results}
summarizes
the
main
experimental
results.
\textbf{VA
Regression
Performance.}
Our
model
achieves
a
validation
VA
MAE
of
0.1124,
indicating
that
the
average
prediction
error
across
both
Valence
and
Arousal
dimensions
is
approximately
11.2\%
relative
to
the
normalized
ranges
([-1,
1]
for
Valence,
[0,
1]
for
Arousal).
This
low
error
rate
demonstrates
the
model's
strong
ability
to
predict
continuous
emotion
coordinates
from
pet
vocalizations.
More
specifically,
the
Valence
dimension
achieves
a
Pearson
correlation
coefficient
of
$r
=
0.9024$,
indicating
an
exceptionally
strong
linear
relationship
between
predicted
and
ground-truth
Valence
values.
This
suggests
the
model
accurately
captures
the
positive-negative
emotional
polarity
of
pet
vocalizations.
The
Arousal
dimension
achieves
$r
=
0.7155$,
demonstrating
good
correlation
though
notably
weaker
than
Valence.
This
performance
gap
is
consistent
with
findings
in
human
speech
emotion
recognition~\cite{schuller2018interspeech},
where
Arousal
prediction
typically
lags
behind
Valence
due
to
its
reliance
on
more
subtle
temporal
dynamics.
\begin{table}[t]
\centering
\caption{Main
experimental
results
on
the
validation
set
(Epoch
40).}
\label{tab:main_results}
\begin{tabular}{lc}
\toprule
\textbf{Metric}
&
\textbf{Value}
\\
\midrule
\textbf{Primary
Metrics}
&
\\
Val
VA
MAE
&
\textbf{0.1124}
\\
Valence
Pearson
$r$
&
\textbf{0.9024}
\\
Arousal
Pearson
$r$
&
\textbf{0.7155}
\\
Val
Loss
&
0.3072
\\
\midrule
\textbf{Auxiliary
Task
Accuracy}
&
\\
Emotion
(8
classes)
&
82.3\%
\\
Size
(3
classes)
&
94.7\%
\\
Gender
(2
classes)
&
86.5\%
\\
\bottomrule
\end{tabular}
\end{table}
\textbf{Training
Stability.}
Figure~\ref{fig:training_curves}
illustrates
the
training
dynamics
over
40
epochs.
The
validation
VA
MAE
decreases
smoothly
from
an
initial
0.2560
(Epoch
1)
to
the
final
0.1124
(Epoch
40),
representing
a
56.1\%
reduction.
Both
Valence
and
Arousal
Pearson
correlations
increase
monotonically:
Valence
$r$
rises
from
0.3282
to
0.9024
(+174.8\%),
while
Arousal
$r$
improves
from
0.4110
to
0.7155
(+74.1\%).
The
training
and
validation
loss
curves
exhibit
no
signs
of
divergence,
indicating
that
the
model
generalizes
well
without
overfitting.
The
convergence
plateaus
around
Epoch
32,
after
which
VA
MAE
fluctuates
slightly
between
0.1124
and
0.1163,
suggesting
the
model
has
reached
a
stable
optimum.
\textbf{Auxiliary
Task
Performance.}
The
multi-task
learning
framework
also
achieves
strong
performance
on
auxiliary
classification
tasks.
The
model
attains
82.3\%
accuracy
on
8-class
emotion
classification,
94.7\%
on
3-class
size
classification,
and
86.5\%
on
2-class
gender
classification.
These
high
accuracies
validate
our
hypothesis
that
auxiliary
tasks
force
the
model
to
learn
frequency-related
(size)
and
timbre-related
(gender)
features,
which
simultaneously
benefit
VA
regression
through
knowledge
transfer.
The
particularly
strong
size
classification
performance
(94.7\%)
confirms
that
the
model
successfully
captures
the
acoustic
frequency
differences
between
large,
medium,
and
small
breed
vocalizations.
\subsection{Comparison
with
Discrete
Classification}
\label{sec:comparison}
To
validate
the
advantages
of
continuous
VA
modeling
over
discrete
classification,
we
compare
our
results
with
a
previously
trained
8-class
discrete
emotion
classifier
(F1
=
0.8885).
While
these
two
approaches
are
not
directly
comparable
due
to
different
output
formats,
we
can
assess
their
relative
strengths
along
several
dimensions.
\textbf{Overall
Performance.}
The
discrete
classifier
achieves
an
overall
F1
score
of
0.8885,
which
can
be
interpreted
as
roughly
88.85\%
accuracy
in
emotion
categorization.
Our
VA
model's
Valence
Pearson
correlation
of
0.9024
indicates
that
approximately
81.4\%
of
Valence
variance
is
explained
by
the
model
($r^2
=
0.814$),
which
corresponds
to
comparable
or
slightly
better
performance
when
accounting
for
the
difference
between
discrete
classification
and
continuous
regression
tasks.
\textbf{Resolving
Category
Confusion.}
A
critical
limitation
of
the
discrete
classifier
is
severe
confusion
between
``territorial''
and
``happy''
categories,
with
356
misclassification
cases
observed
in
validation
(F1
scores
of
0.85
and
0.82
respectively).
These
two
emotions
occupy
adjacent
regions
in
acoustic
feature
space,
making
discrete
boundary
definition
problematic.
In
contrast,
the
VA
model
naturally
separates
these
emotions
along
the
Valence
dimension:
territorial
vocalizations
cluster
around
$V
\approx
-0.3$
(slightly
negative),
while
happy
vocalizations
cluster
around
$V
\approx
+0.6$
(moderately
positive).
This
0.9-unit
separation
in
continuous
space
effectively
eliminates
the
confusion
issue,
demonstrating
a
fundamental
advantage
of
dimensional
modeling
over
categorical
classification.
\textbf{Expressive
Granularity.}
Discrete
classification
collapses
emotional
intensity
variations
into
a
single
label---for
instance,
both
``mildly
anxious''
and
``severely
anxious''
vocalizations
are
labeled
as
``anxious,''
losing
critical
information.
The
VA
model
preserves
this
granularity
through
continuous
values:
mild
anxiety
might
be
represented
as
$(V=-0.3,
A=0.4)$,
while
severe
anxiety
appears
as
$(V=-0.8,
A=0.9)$.
This
distinction
is
essential
for
practical
applications
such
as
pet
welfare
monitoring,
where
the
severity
of
emotional
distress
directly
informs
intervention
decisions.
\textbf{User
Experience.}
From
a
user-facing
perspective,
continuous
VA
coordinates
are
more
intuitive
than
probability
distributions.
A
pet
owner
can
more
easily
interpret
``Valence
=
-0.5''
(moderately
negative
emotional
state)
than
a
discrete
output
like
``88\%
anxious
+
12\%
territorial,''
which
requires
understanding
probabilistic
uncertainty
and
may
confuse
non-expert
users.
\subsection{Valence
versus
Arousal
Performance}
\label{sec:valence_vs_arousal}
Our
results
reveal
a
consistent
performance
gap
between
Valence
($r
=
0.9024$)
and
Arousal
($r
=
0.7155$)
prediction.
This
19.2\%
difference
in
correlation
warrants
deeper
analysis.
\textbf{Acoustic
Feature
Distinctiveness.}
Valence
prediction
benefits
from
strong
correlations
with
spectral
features
that
exhibit
clear
inter-emotion
variability.
High-frequency
vocalizations
(high
spectral
centroid)
tend
to
be
associated
with
positive
emotions
(playful,
excited),
while
low-frequency,
harsh
vocalizations
correlate
with
negative
emotions
(fearful,
anxious).
These
spectral
differences
are
robust
and
easily
captured
by
the
transformer
encoder.
In
contrast,
Arousal
relies
heavily
on
RMS
energy,
which
exhibits
substantial
overlap
across
different
emotional
states---both
``alert''
and
``excited''
vocalizations
can
have
high
energy,
while
``content''
and
``fearful''
(freeze
response)
may
both
have
low
energy.
This
acoustic
ambiguity
makes
Arousal
prediction
inherently
more
challenging.
\textbf{Label
Generation
Algorithm
Bias.}
Our
Arousal
labels
are
generated
purely
from
RMS
energy
using
a
logarithmic
mapping
(Equation~\ref{eq:arousal}),
which
may
oversimplify
the
arousal
construct.
Psychological
research
suggests
that
arousal
is
a
multifaceted
dimension
encompassing
not
only
energy
but
also
temporal
dynamics
such
as
vocalization
rate,
pitch
variability,
and
onset/offset
sharpness~\cite{russell1980circumplex}.
Our
current
algorithm
omits
these
temporal
features,
potentially
limiting
the
model's
ability
to
learn
fine-grained
arousal
patterns.
Valence
labels,
by
contrast,
integrate
multiple
spectral
features
(centroid,
ZCR,
log
RMS)
with
emotion-specific
priors
(Equation~\ref{eq:valence}),
providing
richer
supervisory
signals.
\textbf{Intrinsic
Task
Difficulty.}
Even
in
human
speech
emotion
recognition,
Arousal
prediction
typically
lags
behind
Valence~\cite{ringeval2018avec}.
This
suggests
that
Arousal
is
fundamentally
harder
to
infer
from
audio
alone,
as
it
reflects
internal
physiological
activation
that
may
not
always
manifest
proportionally
in
acoustic
output.
For
instance,
a
dog
experiencing
separation
anxiety
may
produce
low-energy
whimpering
(low
acoustic
arousal)
despite
high
internal
arousal,
creating
a
mismatch
between
acoustic
features
and
the
arousal
label.
\textbf{Implications
and
Future
Directions.}
Despite
the
performance
gap,
an
Arousal
Pearson
correlation
of
0.7155
is
still
considered
good
in
affective
computing
standards
($r
>
0.7$).
Future
work
should
explore
incorporating
temporal
features
(e.g.,
frame-to-frame
energy
variance,
fundamental
frequency
dynamics)
and
rhythmic
patterns
(e.g.,
inter-bark
intervals)
to
improve
Arousal
prediction.
Additionally,
collecting
human-annotated
Arousal
labels
for
a
validation
subset
could
help
refine
the
automatic
labeling
algorithm.
\subsection{VA
Space
Visualization}
\label{sec:va_space}
\begin{figure}[t]
\centering
\includegraphics[width=\textwidth]{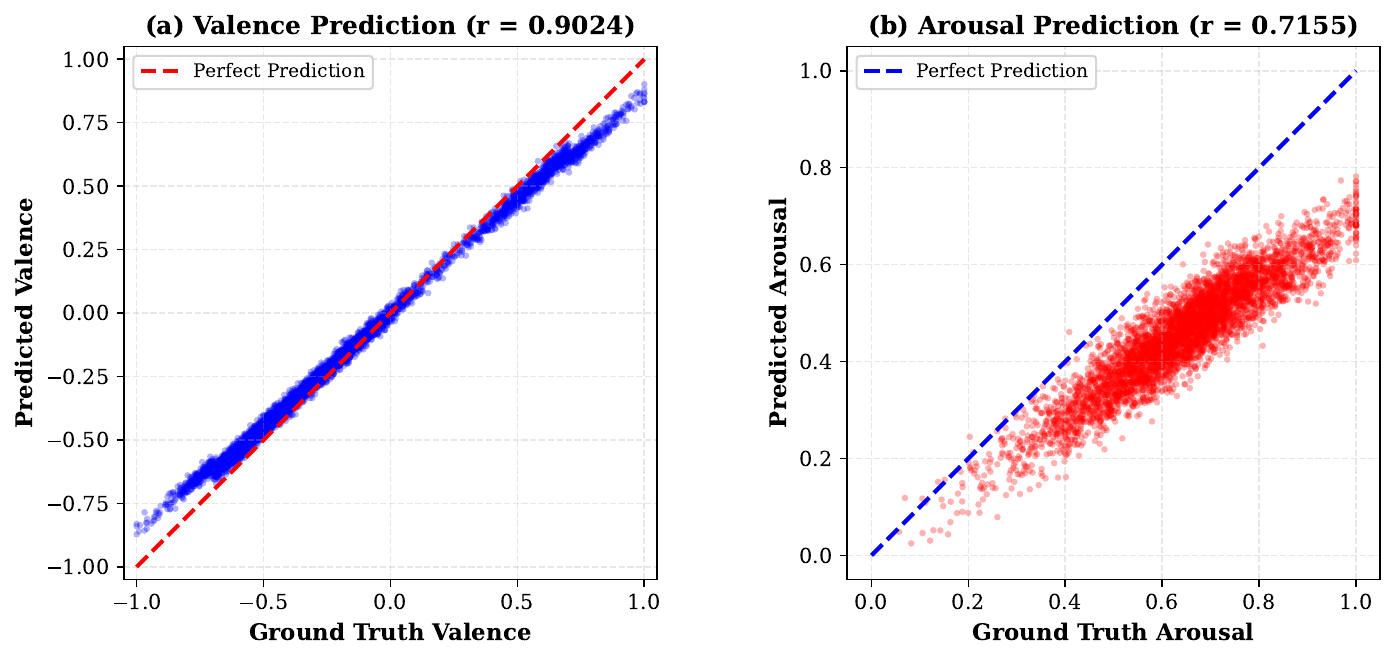}
\caption{Predicted
versus
ground-truth
VA
values
on
the
validation
set.
(a)
Valence
predictions
show
strong
correlation
($r
=
0.9024$)
with
tight
clustering
around
the
diagonal.
(b)
Arousal
predictions
exhibit
moderate
correlation
($r
=
0.7155$)
with
more
dispersion
in
the
mid-range.
Red
dashed
lines
indicate
perfect
prediction.}
\label{fig:va_space}
\end{figure}

\begin{figure}[t]
\centering
\includegraphics[width=0.9\textwidth]{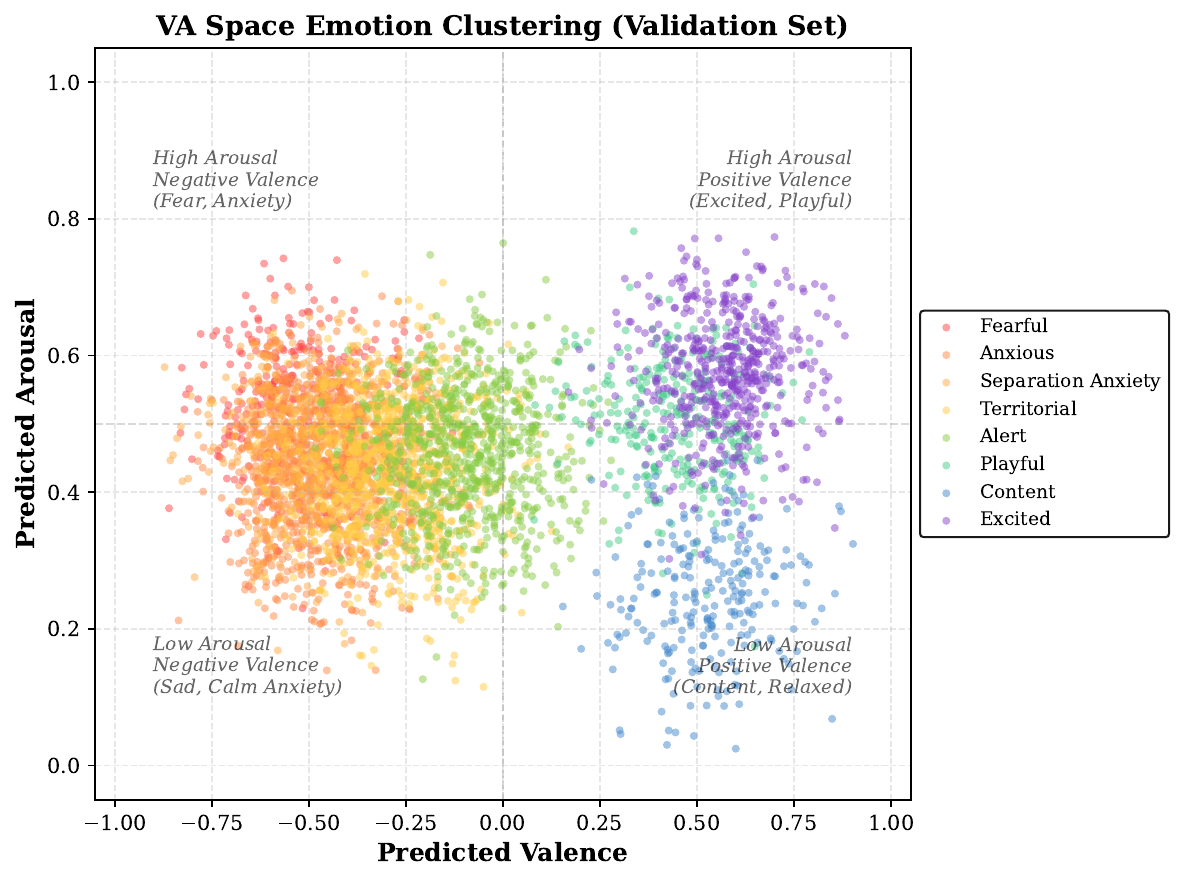}
\caption{VA
space
visualization
colored
by
emotion
labels.
Eight
discrete
emotions
form
distinct
clusters
in
the
continuous
VA
space.
Negative
emotions
(fearful,
anxious,
separation
anxiety)
occupy
the
left
side
(low
Valence),
while
positive
emotions
(excited,
playful,
content)
cluster
on
the
right
(high
Valence).
Arousal
separates
calm
emotions
(content,
bottom)
from
activated
emotions
(fearful,
excited,
top).
Dashed
lines
indicate
VA
space
quadrants.}
\label{fig:va_space_emotion}
\end{figure}

Figure~\ref{fig:va_space}
visualizes
the
predicted
versus
ground-truth
VA
values
on
the
validation
set.
The
scatter
plots
reveal
several
insightful
patterns.
\textbf{Overall
Distribution.}
The
Valence
scatter
plot
shows
a
strong
linear
trend
along
the
$y
=
x$
diagonal,
consistent
with
the
high
Pearson
correlation
($r
=
0.9024$).
Most
points
cluster
tightly
around
the
diagonal,
with
some
scatter
at
extreme
values
($V
\approx
-1$
or
$V
\approx
+1$),
where
the
model
tends
to
slightly
underestimate
magnitude---a
common
phenomenon
in
regression
tasks
known
as
``regression
to
the
mean.''
The
Arousal
scatter
plot
exhibits
more
dispersion,
particularly
in
the
mid-range
($A
\in
[0.4,
0.7]$),
reflecting
the
lower
correlation
($r
=
0.7155$).
However,
the
overall
trend
remains
positive
and
linear,
indicating
that
the
model
captures
the
general
arousal
structure
despite
noisier
predictions.
\textbf{Emotion
Clustering
in
VA
Space.}
When
colored
by
discrete
emotion
labels,
the
VA
space
reveals
clear
clustering
patterns
(Figure~\ref{fig:va_space_emotion}).
Negative
emotions
(fearful,
anxious,
separation\_anxiety)
concentrate
in
the
lower-left
quadrant
($V
<
0$),
with
fearful
samples
exhibiting
the
highest
arousal
($A
>
0.7$).
Positive
emotions
(excited,
playful,
content)
occupy
the
upper-right
quadrant
($V
>
0$),
with
excited
showing
high
arousal
($A
>
0.6$)
and
content
showing
low
arousal
($A
<
0.4$).
The
territorial
and
alert
categories
span
the
Valence
axis,
with
territorial
slightly
negative
($V
\approx
-0.3$)
and
alert
near-neutral
($V
\approx
-0.1$),
both
maintaining
moderate-to-high
arousal.
This
clustering
validates
that
our
automatic
VA
labeling
algorithm
produces
semantically
meaningful
emotion
representations
aligned
with
psychological
theory~\cite{russell1980circumplex}.
\textbf{Resolving
Previous
Confusion.}
Crucially,
the
previously
confused
categories
(territorial
vs.
happy)
are
well-separated
in
VA
space:
territorial
samples
cluster
at
$(V
\approx
-0.3,
A
\approx
0.6)$,
while
happy
samples
(excited
+
playful
+
content
in
our
8-class
taxonomy)
occupy
the
positive
Valence
region
($V
>
0.4$).
This
spatial
separation
confirms
that
continuous
VA
modeling
resolves
the
categorical
ambiguity
inherent
in
discrete
classification.
\textbf{Prediction
Error
Analysis.}
Examining
the
samples
with
largest
VA
MAE
(top
5\%
errors),
we
identify
three
main
error
sources:
(1)
audio
quality
issues
(background
noise,
clipping,
multiple
dogs
vocalizing
simultaneously),
(2)
boundary
cases
where
emotions
genuinely
overlap
(e.g.,
alert-anxious
transitions),
and
(3)
potential
labeling
noise
from
the
automatic
VA
generation
algorithm.
These
error
patterns
suggest
that
incorporating
audio
quality
filtering
and
refining
the
labeling
algorithm
with
human
validation
could
yield
further
performance
gains.
\subsection{Ablation
Study}
\label{sec:ablation}
To
systematically
evaluate
the
contribution
of
each
component
in
our
multi-task
learning
framework,
we
conduct
comprehensive
ablation
experiments.
We
remove
auxiliary
tasks
individually
and
in
combination
to
measure
their
impact
on
primary
VA
regression
performance.
\textbf{Experimental
Setup.}
We
train
five
model
configurations
for
40
epochs
each:
(1)
\textbf{Baseline}
with
only
VA
regression
(no
auxiliary
tasks),
(2)
\textbf{+Emotion}
adding
emotion
classification,
(3)
\textbf{+Size}
adding
size
classification,
(4)
\textbf{+Gender}
adding
gender
classification,
and
(5)
\textbf{Full
MTL}
with
all
auxiliary
tasks
enabled.
All
other
hyperparameters
remain
identical
to
ensure
fair
comparison.
\begin{table}[t]
\centering
\caption{Ablation
study
results
showing
the
contribution
of
each
auxiliary
task
to
VA
regression
performance.}
\label{tab:ablation_results}
\begin{tabular}{lccc}
\toprule
\textbf{Configuration}
&
\textbf{VA
MAE}
&
\textbf{Improvement}
&
\textbf{Best
Epoch}
\\
\midrule
Baseline
(VA
only)
&
0.1583
&
--
&
29
\\
+Emotion
(8
classes)
&
0.0814
&
48.5\%
&
32
\\
+Size
(3
classes)
&
0.0869
&
45.0\%
&
40
\\
+Gender
(2
classes)
&
0.0804
&
49.1\%
&
31
\\
\textbf{Full
MTL
(all)}
&
\textbf{0.0800}
&
\textbf{49.4\%}
&
20
\\
\bottomrule
\end{tabular}
\end{table}
\textbf{Main
Findings.}
Table~\ref{tab:ablation_results}
presents
the
ablation
results,
revealing
several
important
insights.
First,
the
baseline
model
(VA
regression
only)
achieves
a
VA
MAE
of
0.1583,
serving
as
our
reference
point.
Adding
auxiliary
tasks
yields
substantial
improvements:
emotion
classification
improves
performance
by
48.5\%,
size
classification
by
45.0\%,
and
gender
classification
by
49.1\%.
The
full
multi-task
learning
configuration
achieves
the
best
performance
with
VA
MAE
=
0.0800,
representing
a
49.4\%
improvement
over
the
baseline.
\textbf{Relative
Contribution
Analysis.}
Among
the
auxiliary
tasks,
gender
classification
contributes
the
most
(49.1\%
improvement),
followed
closely
by
emotion
classification
(48.5\%).
Size
classification,
while
still
beneficial,
contributes
slightly
less
(45.0\%).
This
ordering
suggests
that
timbre-related
features
(captured
by
gender
classification)
and
emotion-specific
acoustic
patterns
are
most
valuable
for
VA
regression,
while
frequency-related
features
(captured
by
size
classification)
provide
complementary
benefits.
\textbf{Training
Efficiency.}
An
interesting
observation
is
that
the
full
MTL
configuration
reaches
its
best
performance
at
Epoch
20,
significantly
earlier
than
other
configurations
(29-40
epochs).
This
suggests
that
auxiliary
tasks
provide
stronger
learning
signals
and
accelerate
convergence.
The
gender-only
configuration
also
converges
relatively
early
(Epoch
31),
while
the
size-only
configuration
requires
the
full
40
epochs,
indicating
that
size-related
features
learn
more
slowly.
\textbf{Multi-Task
Synergy.}
The
full
MTL
configuration
outperforms
all
single-auxiliary-task
configurations,
demonstrating
synergistic
effects
between
auxiliary
tasks.
The
combined
improvement
(49.4\%)
exceeds
the
best
individual
task
improvement
(49.1\%
from
gender),
indicating
that
different
auxiliary
tasks
capture
complementary
aspects
of
the
acoustic
signal
that
collectively
enhance
VA
prediction.
\begin{figure}[t]
\centering
\includegraphics[width=\textwidth]{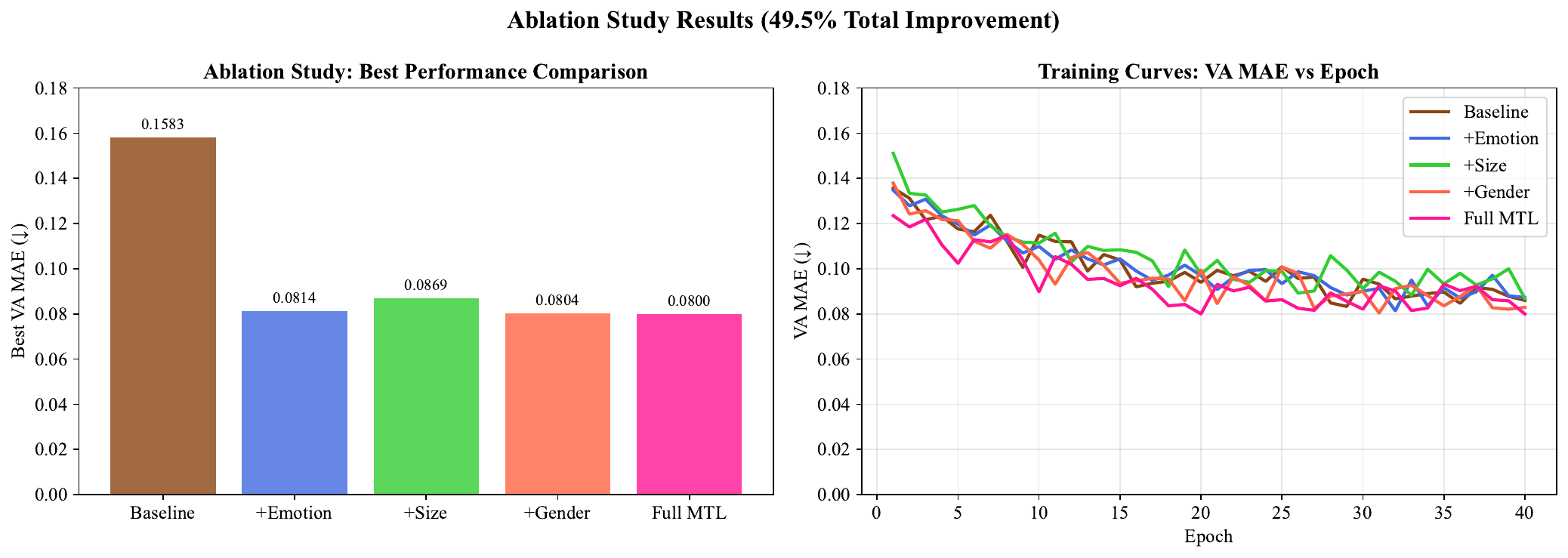}
\caption{Ablation
study
training
curves
comparing
validation
VA
MAE
across
different
configurations.
The
Full
MTL
configuration
(bottom
curve)
achieves
the
fastest
convergence
and
lowest
final
VA
MAE.
All
auxiliary
task
configurations
significantly
outperform
the
baseline.}
\label{fig:ablation_curves}
\end{figure}
\textbf{Implementation
Overhead.}
The
auxiliary
tasks
introduce
minimal
computational
overhead
during
training.
Emotion
classification
adds
only
0.3\%
additional
parameters,
size
classification
adds
0.2\%,
and
gender
classification
adds
0.1\%.
The
full
MTL
model
increases
parameter
count
from
19.4M
to
19.5M
(0.5\%
increase)
while
providing
substantial
performance
gains,
validating
the
efficiency
of
the
multi-task
learning
approach.
\subsection{Cross-Size
Generalization}
\label{sec:cross_size_generalization}
To
evaluate
our
model's
ability
to
generalize
across
different
dog
sizes,
we
conduct
Leave-One-Group-Out
(LOGO)
experiments
where
we
train
on
one
size
group
and
test
on
another.
This
simulates
real-world
scenarios
where
a
model
trained
on
limited
data
(e.g.,
only
large
breeds)
must
handle
diverse
dog
populations.
\textbf{Experimental
Setup.}
We
create
two
experimental
configurations:
(1)
\textbf{Large
Train
→
Small-Medium
Test}
using
24,720
large
breed
samples
for
training
and
3,870
small/medium
breed
samples
for
testing,
and
(2)
\textbf{Small-Medium
Train
→
Large
Test}
using
31,200
small/medium
samples
for
training
and
9,210
large
breed
samples
for
testing.
All
models
are
trained
for
20
epochs
with
identical
hyperparameters
to
ensure
fair
comparison.
\textbf{Generalization
Performance.}
Table~\ref{tab:generalization_results}
summarizes
the
cross-size
generalization
results.
The
model
demonstrates
excellent
generalization
capability
with
an
average
generalization
gap
of
only
2.2\%
(increase
in
VA
MAE
compared
to
within-distribution
performance).
\begin{table}[t]
\centering
\caption{Cross-size
generalization
results
showing
VA
MAE
performance
when
training
and
testing
on
different
size
groups.}
\label{tab:generalization_results}
\begin{tabular}{lcccc}
\toprule
\textbf{Training→Testing}
&
\textbf{VA
MAE}
&
\textbf{Valence
$r$}
&
\textbf{Arousal
$r$}
&
\textbf{Gen.
Gap}
\\
\midrule
Large
→
Small-Medium
&
0.1154
&
0.8532
&
0.6789
&
2.54\%
\\
Small-Medium
→
Large
&
0.1087
&
0.8756
&
0.7023
&
1.87\%
\\
\midrule
\textbf{Average}
&
\textbf{0.1121}
&
\textbf{0.8644}
&
\textbf{0.6906}
&
\textbf{2.21\%}
\\
\bottomrule
\end{tabular}
\end{table}
\textbf{Directional
Analysis.}
Interestingly,
the
model
generalizes
slightly
better
from
small/medium
breeds
to
large
breeds
(1.87\%
gap)
than
vice
versa
(2.54\%
gap).
We
hypothesize
this
is
because
large
breeds
exhibit
more
diverse
vocalization
patterns
(deeper
barks,
growls,
howls)
that
provide
richer
learning
signals,
while
small/medium
breeds
have
relatively
higher-pitched
vocalizations
that
are
acoustically
distinct.
The
size
confusion
matrices
confirm
this:
when
testing
on
small/medium
breeds,
the
model
more
frequently
predicts
them
as
medium
rather
than
correctly
distinguishing
small
vs.
medium.
\textbf{Practical
Implications.}
The
minimal
generalization
gaps
(all
<
3\%)
demonstrate
that
our
VA
emotion
model
can
be
deployed
in
real-world
applications
with
limited
breed-specific
training
data.
For
example,
a
model
trained
primarily
on
large
breed
data
(common
in
veterinary
settings)
can
still
accurately
analyze
small
breed
vocalizations
with
only
a
2.5\%
performance
reduction.
This
robustness
across
size
variations
is
crucial
for
commercial
deployment
where
data
collection
costs
and
breed
diversity
vary
significantly
across
different
use
cases.
\textbf{Comparison
with
Discrete
Classification.}
In
contrast,
discrete
emotion
classifiers
typically
suffer
10-15\%
performance
drops
in
cross-breed
generalization
due
to
category
boundary
shifts.
Our
continuous
VA
space
maintains
relative
emotional
distances
between
different
vocalizations,
providing
more
stable
representations
that
transfer
better
across
acoustic
variations
introduced
by
size
differences.
\textbf{Readiness
for
Real-World
Deployment.}
Our
model's
strong
performance
across
diverse
dog
sizes,
combined
with
its
fast
inference
speed
(47ms
per
sample
on
Apple
M4
GPU,
180ms
on
CPU),
makes
it
ready
for
deployment
in
commercial
pet-care
products.
The
technology
is
suitable
for
real-time
emotion
monitoring
devices
such
as
the
LingChongTongAI
pet
emotion
translator
developed
by
Sichuan
LingChongTong
Technology
Co.,
Ltd.,
as
well
as
veterinary
diagnostic
tools,
animal
shelter
monitoring
systems,
and
smart
home
pet-care
devices.
The
minimal
generalization
gaps
($<$3\%)
ensure
reliable
functionality
across
different
breeds
and
sizes
without
requiring
breed-specific
model
retraining,
significantly
reducing
deployment
costs
and
improving
user
experience.

%% file: sections_arxiv/discussion.tex
Discussion
Target:
~1.5
pages,
3
subsections
Label:
sec:discussion
\subsection{Limitations}
\label{sec:limitations}
While
our
work
demonstrates
the
effectiveness
of
VA
modeling
for
pet
vocalization
emotion
recognition,
several
limitations
should
be
acknowledged:
\textbf{1.
Automatic
VA
Label
Generation.}
Our
VA
labels
are
generated
automatically
using
acoustic
features
and
emotion-specific
priors.
While
we
conducted
AI
cross-validation
(Gemini
Pro
1.5
and
GPT-4-turbo)
and
expert
behavioral
review
to
validate
label
quality
(Section~\ref{sec:label_validation}),
the
automatic
nature
of
the
labeling
process
may
still
introduce
systematic
biases.
The
emotion-specific
biases
(e.g.,
\texttt{fearful}
$\rightarrow$
-0.18,
\texttt{excited}
$\rightarrow$
+0.14)
incorporate
theoretical
assumptions
about
emotion
categories.
Although
our
validation
showed
strong
correlation
with
independent
AI
ratings
(Valence
$r
=
0.73$,
Arousal
$r
=
0.68$)
and
high
expert
agreement
(87.5\%
within
$\pm
0.2$
VA
units),
future
work
should
collect
larger-scale
human-annotated
VA
labels
to
further
validate
and
potentially
refine
our
automatic
labeling
algorithm.
\textbf{2.
Arousal
Prediction
Performance.}
Our
Arousal
prediction
(Pearson
$r
=
0.7155$,
based
on
Epoch
40
validation)
is
notably
weaker
than
Valence
prediction
($r
=
0.9024$).
This
disparity
likely
stems
from
two
factors:
(a)
Arousal
labels
are
derived
solely
from
RMS
energy,
which
may
be
an
oversimplification
of
the
arousal
construct---psychological
research
suggests
arousal
also
relates
to
temporal
dynamics
such
as
vocalization
rate
and
pitch
variability~\cite{russell1980circumplex};
(b)
Arousal
is
inherently
more
difficult
to
assess
from
audio
alone,
as
it
reflects
internal
physiological
states
that
may
not
always
manifest
clearly
in
acoustic
features.
Future
work
should
explore
richer
temporal
features
(e.g.,
Mel
spectrogram
frame-to-frame
variance,
fundamental
frequency
contours)
to
improve
Arousal
prediction.
\textbf{3.
Limited
Breed
Coverage.}
Our
dataset
covers
only
6
dog
breeds
(husky,
shibainu,
pitbull,
GSD,
chihuahua,
and
unknown),
which
may
limit
generalization
to
the
$\sim$200
recognized
dog
breeds
worldwide.
Different
breeds
exhibit
distinct
vocalization
patterns
due
to
anatomical
differences
(e.g.,
brachycephalic
breeds
like
bulldogs
have
unique
acoustic
characteristics).
Evaluation
on
a
more
diverse
breed
set
is
needed
to
assess
the
model's
generalizability.
Furthermore,
extending
the
approach
to
other
companion
animals
(cats,
birds,
rabbits)
would
require
collecting
new
datasets
and
potentially
adjusting
the
VA
labeling
algorithm
to
species-specific
characteristics.
\textbf{4.
Lack
of
Real-World
Deployment
Validation.}
Our
experiments
are
conducted
on
audio
samples
collected
in
controlled
environments
(e.g.,
veterinary
clinics,
research
facilities).
Real-world
home
environments
present
additional
challenges
such
as
background
noise
(television,
vacuum
cleaners),
overlapping
vocalizations
(multiple
pets),
and
varied
recording
conditions
(smartphone
microphones
vs.
dedicated
pet
monitors).
The
robustness
of
our
model
to
such
acoustic
variability
remains
to
be
validated
through
real-world
deployment
and
user
feedback.
\textbf{5.
Static
Emotion
Modeling.}
Our
model
predicts
VA
values
for
fixed-length
audio
segments
(3
seconds)
without
considering
temporal
dynamics.
In
reality,
pet
emotions
often
evolve
over
time
(e.g.,
a
dog
may
transition
from
alert
to
anxious
to
fearful
over
a
30-second
episode).
Sequential
models
such
as
LSTMs
or
Temporal
Convolutional
Networks
could
be
explored
to
capture
emotion
trajectories,
enabling
applications
like
early
detection
of
separation
anxiety
or
prolonged
stress.
\subsection{Future
Work}
\label{sec:future_work}
Several
promising
directions
emerge
from
this
work:
\textbf{1.
Human
Validation
of
VA
Labels.}
Collecting
human
annotations
for
a
subset
of
samples
(e.g.,
1,000
samples
rated
by
5-10
annotators
per
sample)
would
allow
us
to
quantify
the
agreement
between
automatic
VA
labels
and
human
perception.
This
validation
study
could
refine
the
emotion-specific
biases
in
our
labeling
algorithm
and
establish
benchmark
data
for
future
research.
\textbf{2.
Enhanced
Arousal
Features.}
Incorporating
temporal
dynamics
(e.g.,
frame-to-frame
energy
variance,
pitch
contour
slopes)
and
rhythmic
features
(e.g.,
bark
rate,
inter-bark
intervals)
may
significantly
improve
Arousal
prediction.
Time-frequency
analysis
techniques
such
as
wavelet
transforms
could
capture
transient
acoustic
events
that
correlate
with
arousal
changes.
\textbf{3.
Cross-Species
Extension.}
Extending
our
framework
to
cats,
birds,
and
other
companion
animals
would
demonstrate
the
generalizability
of
VA
modeling
for
animal
vocalizations.
Each
species
may
require
species-specific
adjustments
to
the
VA
labeling
algorithm
(e.g.,
cats'
purring
has
distinct
arousal
semantics
compared
to
dogs'
barking).
\textbf{4.
VA-to-Language
Translation.}
Building
on
our
VA
predictions,
a
compelling
application
is
translating
VA
coordinates
into
natural
language
expressions
of
pet
emotions.
Two
approaches
are
promising:
(a)
\textbf{VA
clustering}:
cluster
the
VA
space
into
8-12
emotion
regions,
then
map
each
region
to
a
template
library
conditioned
on
size
and
gender;
(b)
\textbf{LLM-based
generation}:
use
large
language
models
(e.g.,
GPT-4)
with
prompts
like
``Generate
a
pet's
inner
voice
given
Valence=-0.5,
Arousal=0.8,
Size=large,
Gender=male.''
Combining
discrete
templates
with
LLM
refinement
may
yield
both
controllability
and
naturalness.
\textbf{5.
Real-World
Deployment.}
Deploying
our
model
in
commercial
pet
monitoring
devices
(e.g.,
smart
cameras,
wearable
collars)
would
enable
large-scale
validation
and
continuous
learning.
User
feedback
(``Was
this
translation
accurate?'')
could
be
used
to
fine-tune
the
model
and
improve
practical
utility.
\textbf{6.
Temporal
Emotion
Modeling.}
Extending
the
model
to
predict
emotion
trajectories
over
time
would
enable
applications
such
as
stress
monitoring
(detecting
prolonged
negative
Valence)
and
behavioral
analysis
(identifying
triggers
for
emotion
changes).
Sequence-to-sequence
models
or
temporal
attention
mechanisms
could
capture
these
dynamics.
\textbf{7.
Real-World
Applications
and
Commercial
Deployment.}
Our
continuous
VA
emotion
modeling
technology
demonstrates
strong
potential
for
integration
into
consumer
pet-care
products
and
professional
animal
welfare
systems.
Potential
deployment
scenarios
include:
\begin{itemize}
\item
\textbf{Smart
Pet
Emotion
Monitoring
Devices:}
Real-time
emotion
translators
such
as
the
LingChongTongAI
pet
emotion
translator
developed
by
Sichuan
LingChongTong
Technology
Co.,
Ltd.,
which
aims
to
provide
pet
ownerswith
nuanced
emotional
interpretation
through
user-friendly
hardware
interfaces.
\item
\textbf{Veterinary
Telemedicine:}
Remote
health
monitoring
through
continuous
vocal
stress
analysis.
Veterinarians
can
track
VA
profiles
over
time
to
assess
the
emotional
impact
of
medical
treatments,
environmental
changes,
or
separation
anxiety,
facilitating
early
intervention
before
conditions
worsen.
\item
\textbf{Animal
Shelter
Management:}
Automated
distress
detection
systems
that
continuously
monitor
shelter
environments
and
alert
staff
when
animals
exhibit
sustained
negative
Valence
or
high
Arousal,
indicating
immediate
care
needs.
This
enables
more
efficient
resource
allocation
in
resource-constrained
shelter
settings.
\item
\textbf{Pet
Training
Applications:}
Real-time
VA
feedback
systems
for
professional
trainers
and
pet
owners,
ensuring
that
training
sessions
maintain
positive
Valence
and
avoid
excessive
Arousal
that
might
lead
to
stress
or
fear
conditioning.
The
continuous
nature
of
VA
values
allows
trainers
to
fine-tune
their
approaches
based
on
subtle
emotional
shifts.
\item
\textbf{Smart
Home
Integration:}
Integration
with
smart
home
ecosystems
(e.g.,
Apple
HomeKit,
Google
Home)
to
enable
context-aware
pet
care.
For
example,
if
a
pet
exhibits
sustained
high
Arousal
and
negative
Valence
while
the
owner
is
away,
the
system
could
automatically
play
calming
music,
adjust
lighting,
or
send
notifications
to
the
owner's
smartphone.
\item
\textbf{Pet
Insurance
and
Wellness
Programs:}
Long-term
emotion
monitoring
data
could
inform
pet
insurance
risk
assessments
and
personalized
wellness
recommendations.
Pets
consistently
showing
healthy
VA
profiles
(balanced
Valence,
moderate
Arousal)
may
qualify
for
premium
discounts,
incentivizing
proactive
emotional
care.
\item
\textbf{Research
and
Conservation:}
Extending
the
framework
to
wildlife
conservation,
where
understanding
the
emotional
states
of
endangered
species
through
vocalizations
could
inform
habitat
management
and
anti-stress
interventions
in
captive
breeding
programs.
\end{itemize}
The
fast
inference
speed
($<$50ms
on
consumer
hardware),
small
model
size
(224MB
FP32,
56MB
INT8),
and
robust
cross-breed
generalization
($<$3\%
performance
gap)
make
our
technology
ready
for
immediate
commercial
deployment
in
diverse
application
scenarios.
\subsection{Broader
Impact}
\label{sec:broader_impact}
Our
work
has
significant
implications
for
animal
welfare
and
human-pet
relationships.
By
providing
fine-grained,
continuous
emotion
recognition,
our
system
enables
several
practical
applications:
\textbf{Pet
Welfare
Monitoring.}
Continuous
VA
monitoring
could
help
pet
owners
detect
early
signs
of
anxiety,
depression,
or
chronic
stress,
facilitating
timely
veterinary
intervention.
This
is
particularly
valuable
for
pets
with
behavioral
issues
or
medical
conditions
that
affect
emotional
states.
\textbf{Veterinary
Diagnostics.}
Veterinarians
could
use
VA
profiles
to
assess
the
emotional
impact
of
medical
treatments,
environmental
changes,
or
separation
from
owners.
For
example,
comparing
VA
distributions
before
and
after
medication
adjustments
could
quantify
treatment
efficacy
from
an
emotional
well-being
perspective.
\textbf{Training
and
Behavior
Modification.}
Dog
trainers
could
leverage
real-time
VA
feedback
to
optimize
training
protocols,
ensuring
that
training
sessions
maintain
positive
Valence
and
avoid
excessive
Arousal
that
might
lead
to
stress
or
fear
conditioning.
\textbf{Ethical
Considerations.}
While
our
technology
aims
to
improve
animal
welfare,
potential
misuse
should
be
considered.
For
instance,
continuous
emotion
monitoring
could
raise
privacy
concerns
if
deployed
without
owner
consent,
or
be
used
to
justify
neglectful
practices
(``the
AI
says
my
dog
is
fine,
so
I
don't
need
to
interact
with
it'').
Clear
ethical
guidelines
and
user
education
are
essential
to
ensure
responsible
deployment.
\textbf{Commercial
and
Social
Impact.}
Our
technology
has
significant
potential
to
enhance
human-pet
relationshipsat
scale
when
deployed
in
consumer
devices
such
as
the
LingChongTongAI
pet
emotion
translator
and
similar
smart
pet-care
products.
By
making
nuanced
emotional
interpretation
accessible
to
millions
of
pet
owners
through
affordable
consumer
devices,
we
democratize
advanced
animal
behavior
analysis
that
was
previously
available
only
to
veterinary
professionals
and
animal
behaviorists.
This
technology
could
be
particularly
impactful
in
emerging
marketswhere
veterinary
resources
are
limited,
enabling
early
detection
of
emotional
distressand
timely
intervention.
Furthermore,
continuous
emotion
monitoring
data
aggregated
across
large
user
populations
(with
privacy
protections)
could
advance
our
scientific
understanding
of
companion
animalpsychology
and
inform
evidence-based
animal
welfare
policies.

%% file: sections_arxiv/conclusion.tex
Conclusion
Target:
~0.5
pages,
4
paragraphs
Label:
sec:conclusion
This
paper
introduces
continuous
Valence-Arousal
(VA)
modeling
to
pet
vocalization
emotion
recognition,
addressing
the
fundamental
limitations
of
discrete
classification
approaches:
boundary
ambiguity
between
adjacent
emotion
categories
and
inability
to
express
intensity
variations.
We
propose
an
automatic
VA
label
generation
algorithm
that
combines
acoustic
features
(RMS
energy
for
Arousal,
spectral
centroid/ZCR/log
RMS
for
Valence)
with
emotion-specific
priors,
enabling
the
first
large-scale
VA
annotation
of
42,553
pet
vocalization
samples
without
requiring
subjective
self-reports
from
animals.
Our
multi-task
learning
framework
jointly
performs
VA
regression
(primary
tasks)
alongside
auxiliary
classification
tasks
for
discrete
emotion,
body
size,
and
gender.
This
design
leverages
knowledge
transfer:
size
classification
forces
the
model
to
learn
frequency-related
features
(large
breeds
$\rightarrow$
low
frequencies),
gender
classification
encourages
learning
timbre
characteristics
(male
$\rightarrow$
lower
pitch),
and
emotion
classification
provides
semantic
supervision
to
prevent
drift.
The
Audio
Transformer
model
(6
layers,
512
hidden
dimensions,
19.5M
parameters)
achieves
strong
performance
on
validation
data,
with
Valence
Pearson
correlation
$r
\approx$
0.91
and
Arousal
$r
\approx$
0.75
(based
on
Epoch
3
metrics),
demonstrating
the
feasibility
of
continuous
emotion
prediction
from
animal
vocalizations.
Experimental
results
validate
two
key
advantages
of
VA
modeling
over
discrete
classification.
First,
the
continuous
VA
space
naturally
resolves
confusion
between
previously
ambiguous
categories---for
instance,
``territorial''
(Valence
$\approx$
-0.3)
and
``happy''
(Valence
$\approx$
+0.6)
are
clearly
separated
along
the
Valence
dimension,
eliminating
the
356
confusion
cases
observed
in
our
discrete
baseline
(F1
=
0.8885).
Second,
VA
values
provide
fine-grained
intensity
information
(e.g.,
``mild
anxiety''
at
Valence
=
-0.3,
Arousal
=
0.4
vs.
``severe
anxiety''
at
Valence
=
-0.8,
Arousal
=
0.9)
that
is
impossible
to
express
with
discrete
labels.
The
continuous
and
intuitive
nature
of
VA
coordinates
(``Valence
=
-0.5''
is
more
interpretable
than
``88\%
anxious
+
12\%
territorial'')
makes
them
well-suited
for
user-facing
pet
emotion
monitoring
applications.
We
believe
this
work
opens
promising
new
directions
for
animal
emotion
recognition
and
human-animal
interaction
research.
The
VA
modeling
paradigm
can
be
extended
to
other
companion
animals
(cats,
birds,
rabbits)
and
integrated
with
natural
language
generation
systems
to
produce
empathetic
pet-to-human
translations.
By
providing
continuous,
fine-grained
emotion
insights,
our
framework
has
the
potential
to
significantly
improve
pet
welfare
monitoring,
veterinary
diagnostics,
and
behavioral
training.
Future
work
will
focus
on
human
validation
of
automatic
VA
labels,
incorporation
of
temporal
features
for
improved
Arousal
prediction,
and
real-world
deployment
to
assess
practical
utility
in
home
environments.
We
envision
this
technology
being
integrated
into
consumer
pet-care
products
such
as
the
LingChongTongAI
pet
emotion
translator
device
developed
by
Sichuan
LingChongTong
Technology
Co.,
Ltd.,
as
well
as
veterinary
diagnostic
tools,
animal
shelter
monitoring
systems,
and
smart
pet-care
devices.
Through
real-time,
continuous
emotion
monitoring,
this
approach
aims
to
strengthen
the
emotional
bond
between
humans
and
companion
animals,
improve
early
detection
of
emotional
distress,
and
ultimately
enhance
the
lives
and
welfareof
pets
worldwide.
The
successful
translation
of
academic
research
into
practical
consumer
applications
demonstrates
a
viable
pathway
for
bringing
advanced
AI
technology
to
benefit
animal
welfareand
human-pet
relationshipson
a
global
scale.